\documentclass[floatfix,aip,jap,amsmath,amssymb,reprint]{revtex4-2}

\usepackage{amsfonts}
\usepackage{graphicx}
\usepackage{siunitx}
\usepackage{color}
\usepackage{physics} 
\usepackage{multirow} 
\usepackage[utf8]{inputenc}
\usepackage[T1]{fontenc}
\usepackage[colorlinks=true,citecolor=blue,urlcolor=red]{hyperref}

\newcommand{\xn}{X}

\begin{document}

\title{Silicon cantilevers locally heated from 300$\,$K up to the melting point: temperature profile measurement from their resonances frequency shift}

\author{Basile Pottier}
\affiliation{Univ Lyon, Ens de Lyon, Univ Claude Bernard Lyon 1, CNRS, Laboratoire de Physique, F-69342 Lyon, France}

\author{Felipe Aguilar Sandoval}
\affiliation{Departamento de Ciencias Naturales y Tecnología, Universidad de Aysén, Obispo Vielmo 62, Coyhaique, Chile}

\author{Mickaël Geitner}
\affiliation{Univ Lyon, Ens de Lyon, Univ Claude Bernard Lyon 1, CNRS, Laboratoire de Physique, F-69342 Lyon, France}

\author{Francisco Esteban Melo}
\affiliation{Departamento de Física and Center for Soft Matter Research, SMAT-C, Universidad de Santiago de Chile, Avenida Ecuador 3493, Estación Central, 9170124, Santiago Chile}

\author{Ludovic Bellon}
\email{ludovic.bellon@ens-lyon.fr}
\affiliation{Univ Lyon, Ens de Lyon, Univ Claude Bernard Lyon 1, CNRS, Laboratoire de Physique, F-69342 Lyon, France}

\date{\today}

\begin{abstract}
When heated, micro-resonators present a shift of their resonance frequencies. We study specifically silicon cantilevers heated locally by laser absorption, and evaluate theoretically and experimentally their temperature profile and its interplay with the mechanical resonances. We present a enhanced version of our earlier model [F. Aguilar Sandoval \textit{et al.}, J. Appl. Phys. \textbf{117}, 234503 (2015)] including both elasticity and geometry temperature dependency, showing that the latter can account for 20\% of the observed shift for the first flexural mode. The temperature profile description takes into account thermal clamping conditions, radiation at high temperature, and lower conductivity than bulk silicon due to phonon confinement. Thanks to a space-power equivalence in the heat equation, scanning the heating point along the cantilever directly reveals the temperature profile. Finally, frequency shift measurement can be used to infer the temperature field with a few percent precision.
\end{abstract}

\maketitle 

\section{Introduction}

Time is the most accurate quantity that can be measured. For this reason, many sensors convert a measurand (e.g. temperature) to time or frequency. Another long term trend in sensing is miniaturization, leading to ever more integrated, low power, low cost and reliable detectors. At the overlap of those processes stand mechanical micro-resonators, with countless applications: time itself~\cite{vanBeek-2012}, quantum state detection~\cite{Verhagen-2012}, mass detectors~\cite{Thundat-1994,Dohn-2005}, chemical and biological sensors~\cite{Lavrik-2004,Tamayo-2012}, flow meters~\cite{Barth-2005,Salort-2012}, force sensors~\cite{Mamin-2001,deLepinay-2017}, thin films mechanical characterisation~\cite{Granata-2015,Li-2014}, temperature measurement~\cite{Spassov-1992,Zhang-2020}. The increased sensitivity of those detectors opened the door to single molecule characterisation, as demonstrated for example by imaging their atomic structure~\cite{Gross-2009}, measuring their optical absorption~\cite{Chien-2018} or real-time monitoring of receptor-ligand interactions~\cite{Tamayo-2012}. 

Temperature and related properties stand among the observables accessible with micro-mechanical resonators: as temperature changes, due to heat exchanges between the resonator and its environment, its geometrical and mechanical properties evolve, hence its resonance frequency. Quartz based temperature sensors~\cite{Spassov-1992} have for example been available for half a century. More recently, MEMS (Micro Electro Mechanical Systems) based bolometers have demonstrated high potential for infrared imaging~\cite{Dao-2019} or spectroscopy. Optical absorption can for instance be monitored by shining light on molecules resting of a membrane~\cite{Chien-2018}: absorption leads to heating, temperature elevation, and resonance frequency shift of the membrane. In the same spirit, infrared (IR) absorption of a surface can be tested at the nanoscale by coupling local thermal expansion due to IR absorption to Atomic Force Microscopy (AFM): modulating the IR source at the resonance frequency of the AFM cantilever, unprecedented resolution can be achieved~\cite{Lu-2014}. The photo-thermal effect can also be applied directly to an AFM cantilever is use to drive it at resonance in dynamic measurements~\cite{Marti-1992,Allegrini-1992,Ramos-2006,Kiracofe-2011,Bircher-2013}.
Moreover, the thermo-mechanical coupling has been recently applied to fine tuning of the mechanical response of micro-resonators and nano-photonics devices through localized heating produced by focused light \cite{Jiang2020, Koehler2018, Kumar2013, Keitz2018, Makarov2017}. 

In these examples, coupling between light absorption, heating and the mechanical resonator is the key ingredient and a desirable feature. In other cases, light absorption and heating are just a side-effects of the light used to measure other properties: sensing the deflection of AFM cantilevers with the optical lever technique~\cite{Meyer-1988} or interferometry~\cite{Rugar-1989,Schonenberger-1989,Mulhern-1991,Mamin-2001,Hoogenboom-2008,Paolino-2009-JAP,Laurent-2012,Laurent-2013,Paolino-2013}, measuring Raman Spectra to infer material composition~\cite{Zobeiri2020}, strain~\cite{LiQiu2020} or temperature~\cite{McCarthy-2005,Lee-2006,milner_heating_2010,Chen-2011} for example. In any case, a fine understanding of the thermo-mechanical coupling is important to reach quantitative and robust results.

We focus in the present work on a basic design for the resonator, namely a rectangular cantilever shape. This is especially significant for AFM applications, including scanning thermal microscopy~\cite{gomes_2015}, but the approach could be extended to other application and devices, like membranes or tuning forks. Even when heated cantilevers are not used in a resonant mode, our methods can be of interest to quantify a priori the opto-thermo-mechanical coupling. In a previous work~\cite{aguilar_sandoval_resonance_2015}, we presented experimental evidence of a frequency shift of the resonant modes of a silicon cantilever when the light power of the optical measurement set-up is increased. This frequency shift was identified as the signature of a temperature profile along the cantilever, and a model describing the results was presented: all resonance modes are affected, with frequency shifts values governed by the temperature field, the Young's modulus temperature dependence, and the spatial mode shape. An agreement around 25\% between modes of the temperature elevation was achieved with this framework, turning the frequency shift measurement into an actual temperature sensor for the cantilever.

The main goal of this article is to provide a framework to determine quantitatively the temperature field of a locally heated cantilever, from the measurement of its resonances frequency shift. We refine the analysis with respect to our previous work\cite{aguilar_sandoval_resonance_2015} by including the effect of heat radiation and of thermal expansion of the cantilever in the model. We show that the former effect is noticeable at high temperatures, and the latter is particularly important to consider for the first resonance mode: neglecting it can lead to discrepancies up to 20\% in temperature estimation for a non uniform thermal gradient. This is all the more important as in many experiments, the first oscillation mode is the only measured one: it is the lowest frequency one (and thus the most accessible), and its displacement amplitude is usually larger than other modes, so that the majority of transduction schemes has been primarily developed to detect it. One distinctive feature of our method is actually to include more resonant modes in the analysis, providing a more robust measurement strategy. To back our approach, we present a comprehensive set of new measurements, including Raman spectroscopy for a direct temperature field measurement~\cite{McCarthy-2005,Lee-2006,milner_heating_2010,Chen-2011}, thermal noise~\cite{aguilar_sandoval_resonance_2015} or driven resonance frequency shifts tracking, and temperature profile exploration through heating point scanning. Our measurement eventually reach a quantitative agreement within a few percent between all measured modes and the theoretical picture provided (a tenfold improvement with respect to our previous work~\cite{aguilar_sandoval_resonance_2015}), at all temperatures between ambient and silicon melting.

The article is organized as follows: In section~\ref{sectionThermal}, we determine the temperature profiles for a cantilever heated by a laser in vacuum. We evaluate the effects of thermal radiation and discuss the thermal boundary condition at the clamp. Raman measurements are presented to support our conclusions. In section \ref{ModelFreqShift}, we present an analytical model predicting the resonance frequency of the flexural modes for a non-uniform temperature of the cantilever. We consider both the thermally-induced changes of intrinsic elasticity and geometrical dimensions. We evaluate and compare the sensitivity to these two thermo-mechanical effects for different temperature profiles, in particular for the specific case of silicon cantilever. In section \ref{Experiments}, we apply this framework to measurements in vacuum, and demonstrate the quantitative agreement between observations and theory. As a side-product to this study, we also show that frequency shift measurements alone allow the reconstruction of the temperature profile of the cantilever, thus could give access to the thermal conductivity of its material.

\section{Thermal problem: cantilever heated in vacuum by laser irradiation}\label{sectionThermal}

Heated AFM cantilevers have been extensively studied in the literature~\cite{Marti-1992,Allegrini-1992,Thundat-1994,McCarthy-2005,Ramos-2006,Lee-2006,lee_thermal_2007,milner_heating_2010,Chen-2011,Kiracofe-2011,Bircher-2013,aguilar_sandoval_resonance_2015,gomes_2015}, we gather in this section the main ingredients useful for a quantitative description of the temperature field of a silicon cantilever. We consider a rectangular cantilever of length $L$ much larger than its width $B$, itself much larger its thickness $H$, clamped to a macroscopic chip at temperature $T_0$, and heated by a laser beam focused at some distance $x_0$ from the clamp. In the limit where the cross section dimensions $B,H$ are small compared to $x$ and $L$, the temperature $T$ may be assumed homogeneous across the cross section. $T$ therefore only depends on the longitudinal coordinate $x$ and is described by the one-dimensional heat diffusion equation
\begin{align}\label{HeatEquation}
	\rho c_p\frac{\partial T}{\partial t} - \frac{\partial}{\partial x} \left( \lambda \frac{\partial T}{\partial x} \right)&= q(x),
\end{align}
where $ c_p$ is the heat capacity, $\lambda$ the thermal conductivity and $q$ the heat source/sink density at each position. The characteristic time for heat diffusion along the cantilever is $\rho c_p L^2 /\lambda$, of the order of a few ms for our samples. Since all the measurements will be performed on a much slower time scale, we will consider Eq.~\ref{HeatEquation} in the stationary regime ($\partial T/\partial t=0$).

\subsection{First approach: isothermally clamped edge, no radiation}

\begin{figure}[htb]
 \centering
 \includegraphics[scale=.5]{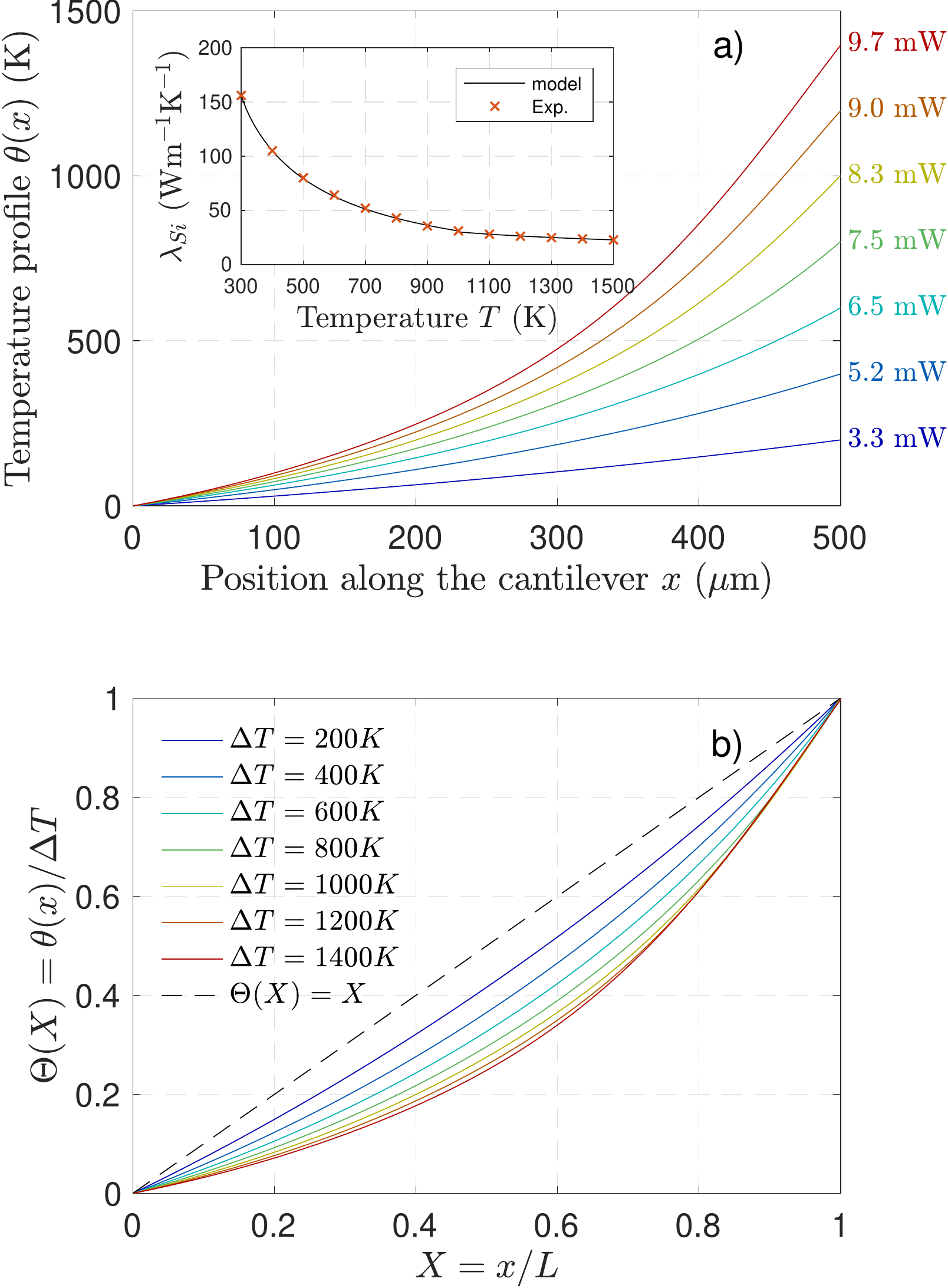}
 \caption{\label{FigThProfileNoRadiation} a) Numerical temperature profiles computed from Eq.~\eqref{IntHeatEquation} for a silicon cantilever in vacuum with $L=\SI{500}{\micro m} $ and $BH=\SI{75}{\micro m^2}$, heated at $x_0=L$ by a constant power $P_a$ labelled on the right of each curve. Inset: temperature dependence of the silicon conductivity $\lambda$ from Refs.~\onlinecite{glassbrenner_thermal_1964, prakash_thermal_1978}. b) Temperatures profiles of panel a) normalized by their respective maximum temperature increase $\Delta T=\theta(x=L)$.}
\end{figure}

As the cantilever is placed in vacuum, no heat transfer can occur by convection with its surroundings. In this first step, we neglect thermal radiation, the only possible heat transfer mechanism is thus thermal conduction through the cantilever. All cantilever surfaces are then assumed to be thermally insulated, except at the heating point $x_0$. The temperature is thus solution of
\begin{equation}\label{HeatEquation0}
	 \dv{}{x}\left( \lambda (T) \dv{T}{x} \right)= \frac{P_a}{B H} \delta_D(x-x_0),
\end{equation}
where $P_a=A P_0$ is the absorbed power (fraction $A(T(x_0))$ of the power $P_0$ of the laser), and $\delta_D$ is Dirac's distribution.
The temperature profile may be obtained by integrating twice \eqref{HeatEquation0} and imposing the boundary condition of an isothermally clamped edge and a thermally insulated free end:

\begin{subequations}
\label{BCThermal}
\begin{align}
	T(x=0)&= T_0 \\
	\left. \frac{ \dd T }{\dd x } \right\vert_{x=L} &= 0.
\end{align}
\end{subequations}
The temperature increase profile $\theta(x)=T(x)-T_0$ is thus solution of 
\begin{subequations}
\label{IntHeatEquation}
\begin{eqnarray}
	 \int_{T_0}^{T_0+\theta(x)} \lambda (T')\dd T' &= \frac{P_a}{B H}x \quad & \mathrm{for} \quad x\le x_0 \\
	 \theta(x)&=\theta(x_0) \quad & \mathrm{for} \quad x> x_0
\end{eqnarray}
\end{subequations}

Using the silicon conductivity data from Ref.~\onlinecite{glassbrenner_thermal_1964} displayed in the inset of figure~\ref{FigThProfileNoRadiation}-a, we can numerically solve Eq.~\eqref{IntHeatEquation} and obtain the temperature profiles $\theta(x) $ for various absorbed power $P_a$ and $x_0=L$. We plot these estimations in figure~\ref{FigThProfileNoRadiation}-a for a cantilever such that $L=\SI{500}{\micro m}$ and $BH=\SI{75}{\micro m^2}$, with $T_0=\SI{22}{\degree C}$. In figure~\ref{FigThProfileNoRadiation}-b we report the corresponding normalized temperature profiles $\Theta(\xn)= \theta(x)/ \Delta T$ with $\Delta T$ the maximum temperature increase and $\xn=x/L$ the normalised position. The non-linear shape of the temperature profiles due to the significant temperature dependency of the silicon conductivity appears clearly.

\begin{figure}[htb]
 \centering
 \includegraphics[scale=.5]{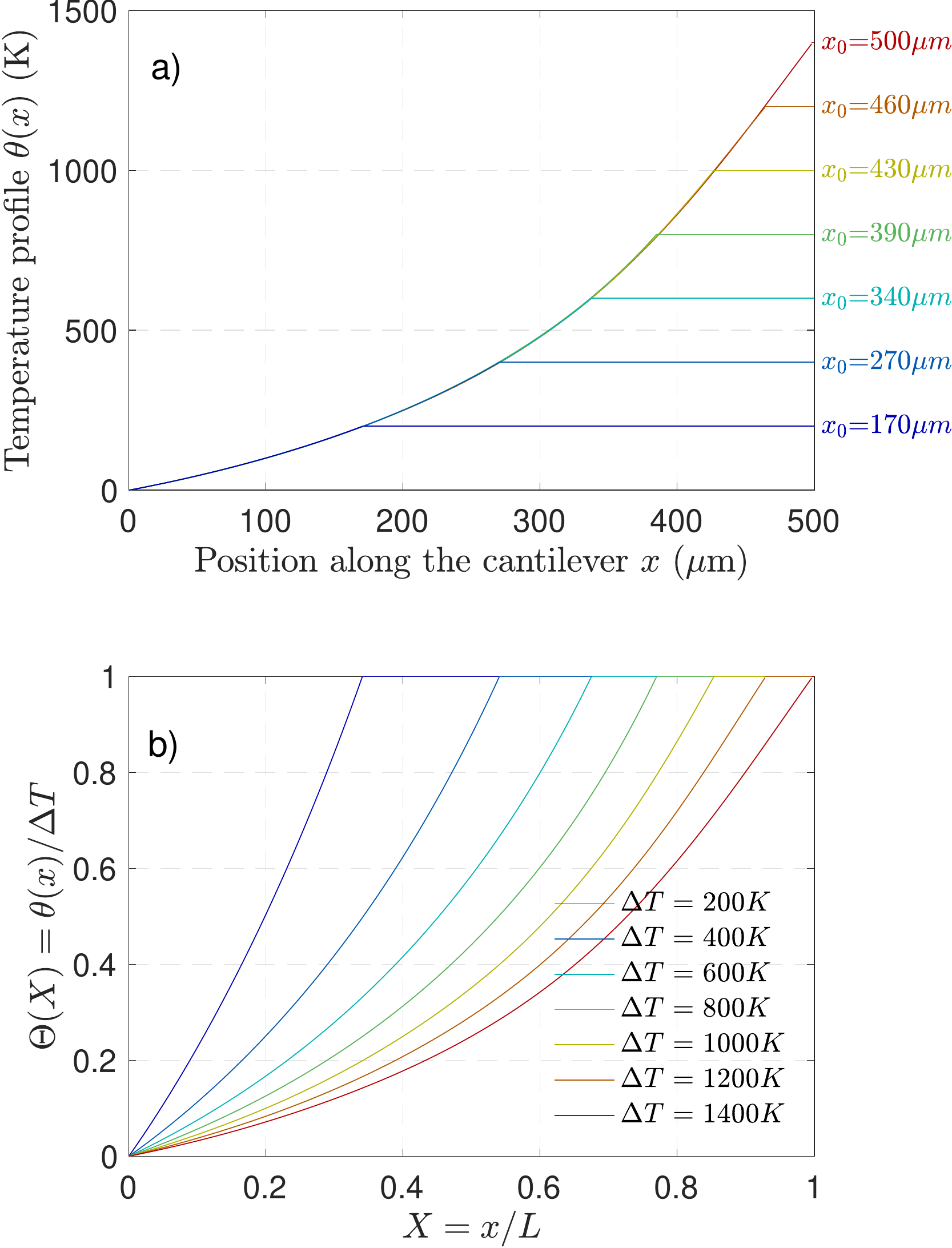}
 \caption{\label{FigThProfileNoRadiationx0} a) Numerical temperature profiles computed from Eq.~\eqref{IntHeatEquation} for a silicon cantilever in vacuum with $L=\SI{500}{\micro m} $ and $BH=\SI{75}{\micro m^2}$, heated by a constant power $P_a=\SI{9.7}{mW}$ at position $x_0$ labelled on the right of each curve. b) Temperatures profiles of panel a) normalized by their respective maximum temperature increase $\Delta T=\theta(x=L)$.}
\end{figure}

When the laser spot is located at $x_0<L$, the temperature profiles are just truncated in their upper part so that $\theta(x>x_0)=\Delta T$, as illustrated in Fig.~\ref{FigThProfileNoRadiationx0}. Therefore, scanning the cantilever with the heating position $x_0$ and reading the temperature at the same position $\theta(x_0)$ will directly draw the temperature profile of Fig.~\ref{FigThProfileNoRadiation}-a.

It is worth noting that in this case where conduction is the only dissipating mechanism, the normalized profiles $\Theta(\xn)$ are independent of the geometry $B,H,L$. The profiles displayed in Fig.~\ref{FigThProfileNoRadiation}-b are fully determined by the intrinsic material conductivity $\lambda(T)$.

\begin{figure}[htb]
 \centering
 \includegraphics{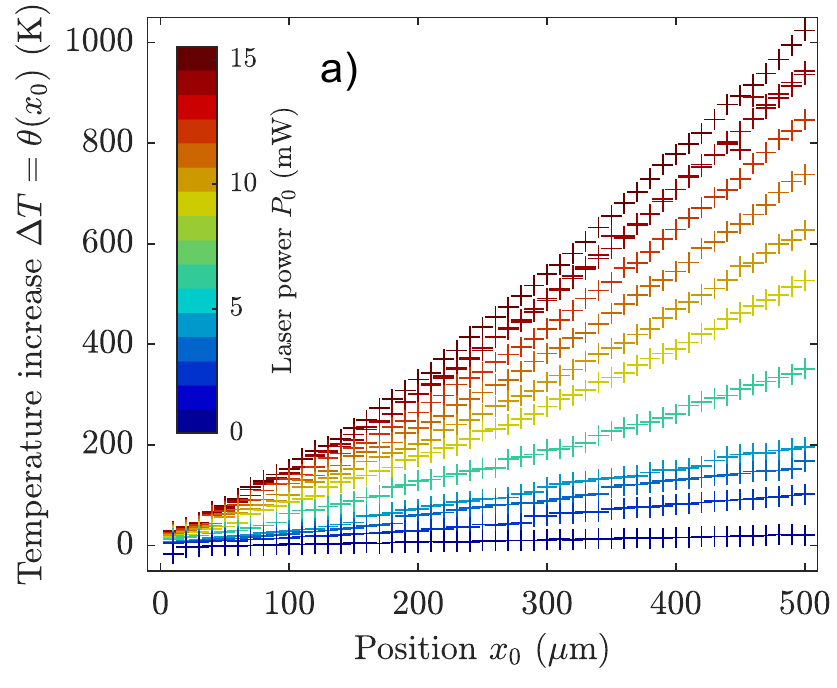}
 \includegraphics{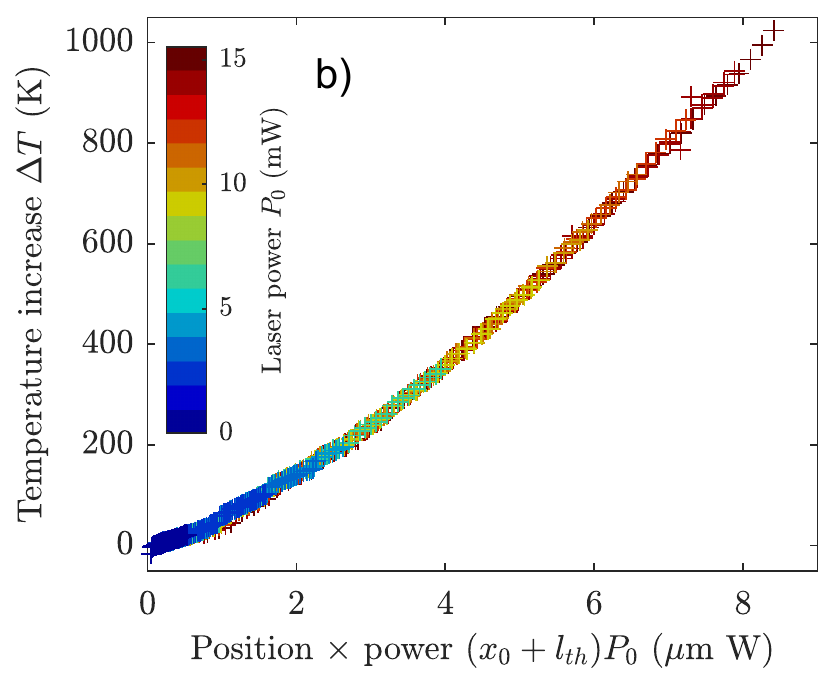}
 \caption{\label{FigRaman} a) Experimental temperature profiles measured with Raman spectroscopy for a BudgetSensor AIO tipless silicon cantilever in vacuum ($L=\SI{500}{\micro m}$, $BH=\SI{78}{\micro m^2}$), as a function of the heating/sensing laser position $x_0$ and power $P_0$. b) Data of panel a) plotted as a function of the product $(x_0+l_{th})P_0$. The offset $l_{th}=\SI{40}{\micro m}$ applied to $x_0$ accounts for the thermal resistance of the cell and chip till the cantilever clamp. All data collapse on a master curve, as expected from eq.~\eqref{EqTempSolveWithLth}.}
\end{figure}

Another characteristic resulting from Eq.~\eqref{HeatEquation0} is the same dependence of the temperature increase to the position $x$ and the absorbed power $P_a$. The variations of the maximum temperature rise $\Delta T$ with the absorbed power $P_a$ corresponds exactly to the variations displayed in figure~\ref{FigThProfileNoRadiation}-a where $P_a$ would linearly vary from zero to the indicated values. To illustrate this behavior, we report in Fig.~\ref{FigRaman} a measurement using a Raman spectrometer to track the temperature of the cantilever while sweeping both $P$ and $x_0$ (experimental details in appendix \ref{AppendixRaman}). Fig.~\ref{FigRaman} shows that up to $\Delta T = \SI{1000}{K}$, the temperature profile depends on $x_0$ and $P_0$ through their product only. Such equivalence is very interesting to evaluate a temperature profile, since scanning in power at a fixed position is generally easier to perform than scanning in space at a fixed power.

\subsection{On the boundary condition of isothermally clamped edge}

\begin{figure}[htb]
 \centering
 \includegraphics[scale=.5]{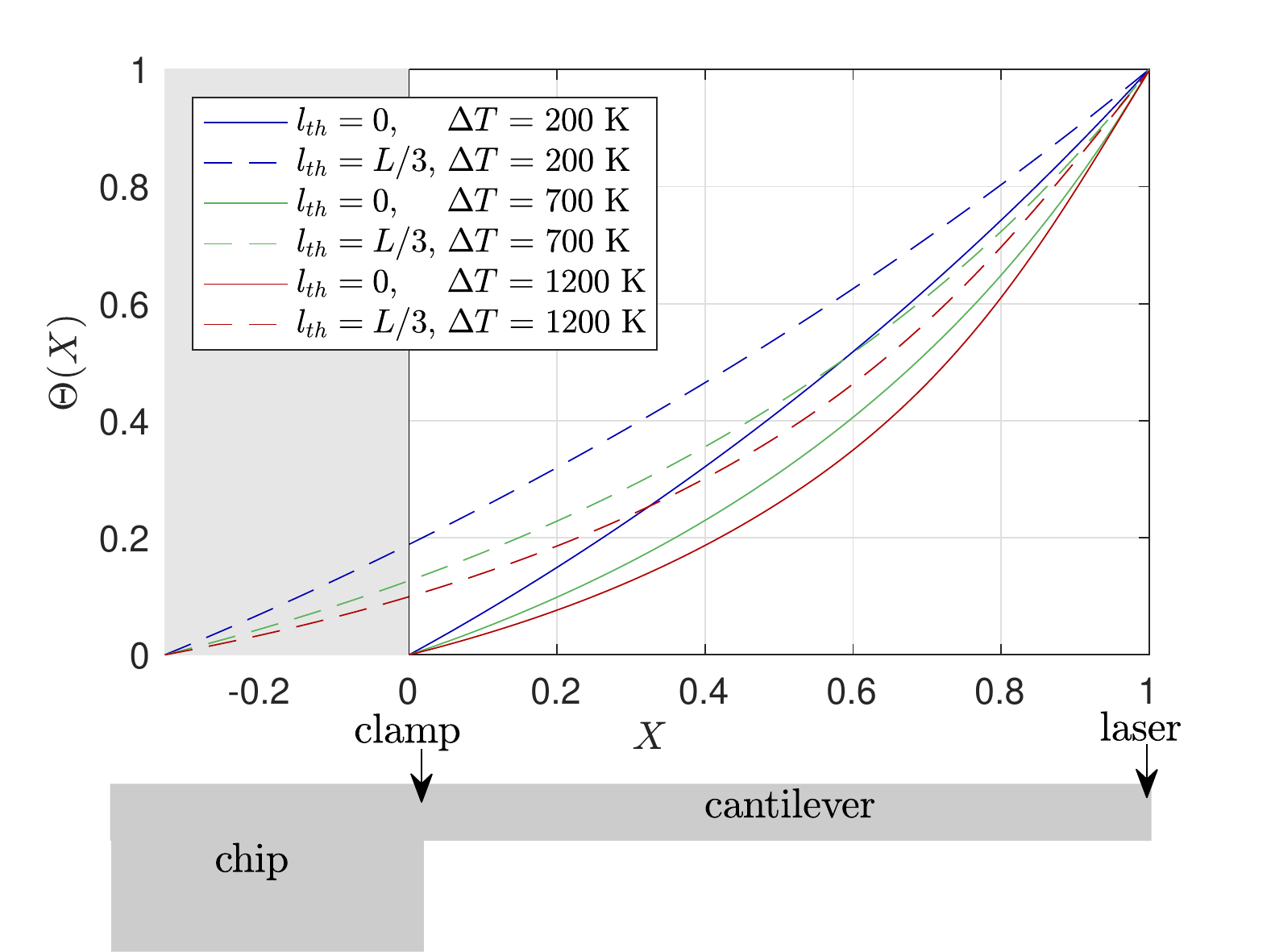}
 \caption{\label{FigBCThermal} Effect of the temperature rise at the clamp on the temperature profiles. The normalized temperature profiles $\Theta(\xn)$ are obtained solving Eq.~\eqref{HeatEquation0} with the boundary conditions $T(x=0)=T_0$ (solid lines) or $T(x=0)=T_\mathrm{clamp}$ such that $l_\mathrm{th}=L/3$ (dash lines). These functions are independent of the geometry. Such large $l_\mathrm{th}$ has been chosen for illustration purposes, but in the experiments $l_\mathrm{th}$ is in the few percent range of $L$.}
\end{figure}

So far we assumed that the temperature increase at the clamp, between the cantilever and the chip supporting it, was negligible, writing the boundary condition $T(x=0)=T_0$. However, no matter how large is the chip compared to the cantilever, most of the heat absorbed from the laser is dissipated inside the chip, inevitably resulting in a rise of the cantilever temperature at the clamp. In this section we evaluate the temperature rise at the frontier $x=0$ considering the dissipation problem inside the chip. In the limit $B\gg H$, we can reduce the problem to two dimensions, the thermal flux $J$ inside the chip is radial and inversely proportional to the distance $r$ from the clamp $J=2P_a/\pi r B$. Hence the temperature decays as $T_\mathrm{chip}(r)=2 P_a/\pi \lambda_0 B \ln\left(l_{\mathrm{chip}}/r\right)+T_0$ where $l_{\mathrm{chip}}$ denotes the distance from the clamp where the temperature is assumed to be equal to the reference temperature $T_0$. Assuming a good thermal contact, $l_{\mathrm{chip}}$ is fixed by the chip geometry between the cantilever base and the thermostat at $T_0$. The temperature at the clamp ($r\rightarrow 0$) can be estimated choosing a cutoff for small $r$ equal to cantilever thickness $H$, 
\begin{equation}\label{Tclamp}
	T_\mathrm{clamp}=\frac{2P_a}{\pi B \lambda_0}\ln \left( \frac{ l_{\mathrm{chip}}}{H}\right) +T_0,
\end{equation}
Using the new boundary condition $T(x=0)=T_\mathrm{clamp}$, the cantilever temperature increase $\theta(x)$ solution of \eqref{HeatEquation0} can be expressed in the similar form than \eqref{IntHeatEquation} 
\begin{align}\label{EqTempSolveWithLth}
	 \int_{T_0}^{T_0+\theta(x)} \lambda_\mathrm{Si}(T')\dd T' &= \frac{ P_a}{B H} (x +l_{\mathrm{th}} ), 
\end{align}
where $l_{\mathrm{th}}$ corresponds to the distance from the clamp where the cantilever temperature would extrapolate to $T_0$ (figure~\ref{FigBCThermal}) and is defined by
\begin{equation}\label{ThermalLength}
	l_\mathrm{th} = \frac{2H}{\pi}\ln\left( \frac{ l_{\mathrm{chip}}}{H} \right).
\end{equation}
The introduction of $l_\mathrm{th}$ allows to take into account the temperature increase at $x=0$. According to eq.~\eqref{ThermalLength} the effect of the temperature increase at the clamp is governed by cantilever thickness $H$. For $H=\SI{2.5}{\micro m}$ and $l_\mathrm{chip}=\SI{300}{\micro m}$ (typical chip thickness value), we get $l_\mathrm{th}=\SI{7.6}{\micro m}$. Performing a 3D simulation with COMSOL Multiphysics for a cantilever with $H=\SI{2.5}{\micro m}$ and $B=\SI{32}{\micro m}$, we find $l_{\mathrm{th}}=\SI{7.8}{\micro m}$. Note that any additional thermal resistance between the chip and the thermostat at temperature $T_0$ will increase the value of $l_{th}$.

In many practical cases, since $l_{th}\ll L$, this refined thermal boundary condition has only a tiny effect on the behavior of the system, and usually goes unnoticed. For example, for a cantilever such that $H=\SI{2.5}{\micro m}$ ($l_\mathrm{th}=\SI{7.6}{\micro m}$) and $L=\SI{500}{\micro m}$, the error neglecting the temperature increase at the clamp is only of $l_\mathrm{th}/L=1.5$~\% on the temperature increase at the free end. However, as we will see in section~\ref{ModelFreqShift}, the frequency shift of the fundamental mechanical resonance is mainly sensitive to the temperature variation close to the clamp. It is thus important to take into account the chip temperature increase (characterized by the length $l_\mathrm{th}$) in order to correctly predict the frequency shift of such mode.

\subsection{Thermal radiation effects}

\begin{figure}[htb]
 \centering
 \includegraphics[scale=.5]{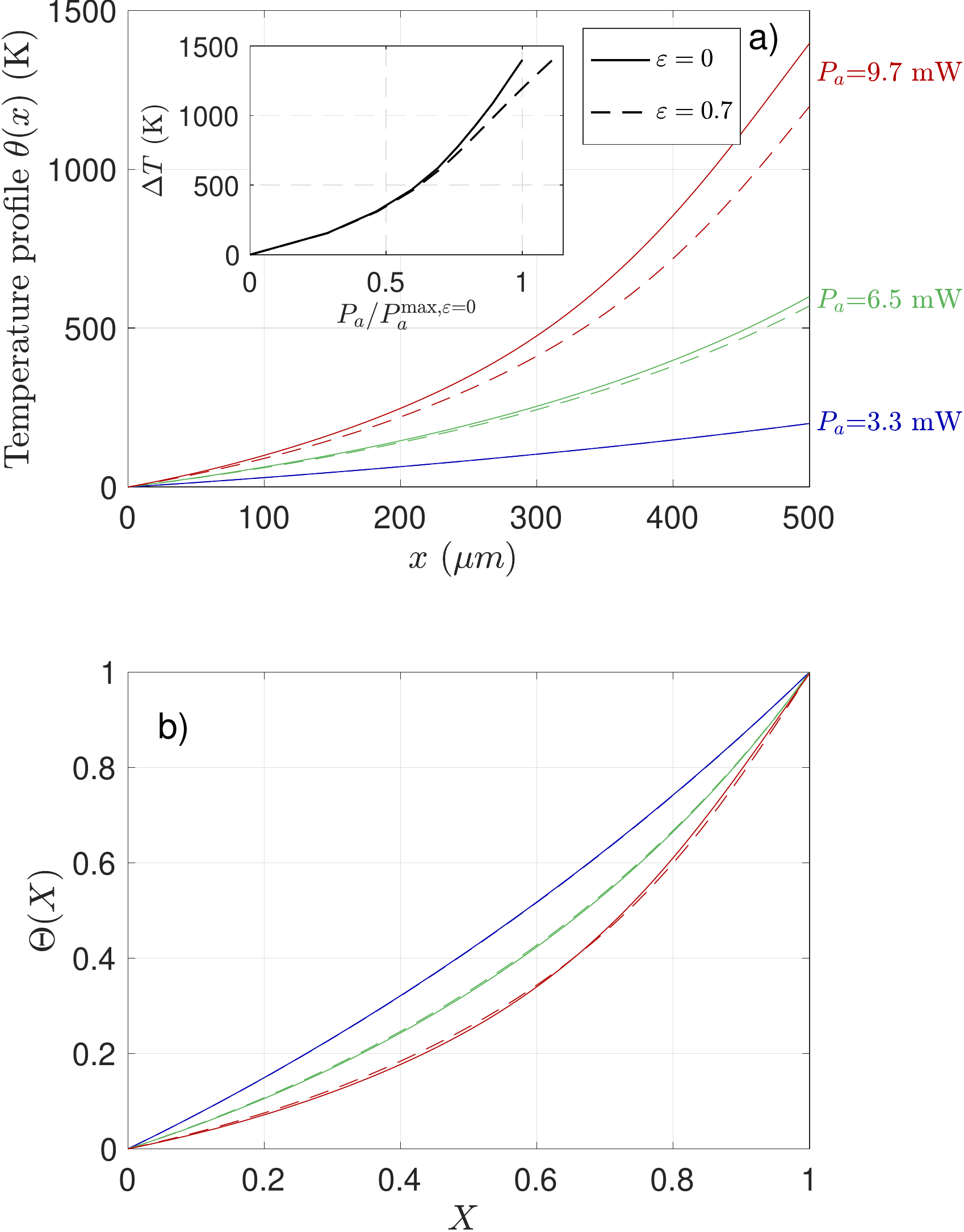}
 \caption{\label{FigThRadiationEffect} a) Numerical temperature profiles with ($\varepsilon=0.7$) and without ($\varepsilon=0$) the radiation heat transfer . The cantilevers dimensions are $H=\SI{2.5}{\micro m}$, $B=\SI{30}{\micro m}$ and $L=\SI{500}{\micro m}$. Inset: Maximum temperature increase $\Delta T$ (at $x=L$) versus the absorbed power $P_a$ normalized by power needed to reach the melting point for a non radiative cantilever $P_a
^{\max, \varepsilon=0}$. b) Temperatures profiles of panel a) normalized by their respective maximum temperature increase $\Delta T=\theta(x=L)$. Thermal radiation has a negligible effect on the functions $\Theta(\xn)$.}
\end{figure}

In the previous paragraphs, the cantilever surfaces were considered adiabatic ($q=0$), neglecting both thermal radiation and convection with the surroundings atmosphere. While placing the cantilever in vacuum enables to neglect the convection, one shall still consider the radiative effects, especially as high temperatures are reached. In general, the output power by a radiating surface $S$ at the temperature $T$ is given by $\varepsilon \sigma S T^4 $, with $\sigma= \SI{5.67e-8}{W.m^{-2}.K^{-4}}$ the Stefan-Boltzmann constant and $\varepsilon$ the material emissivity ($\varepsilon =1$ for a perfect black body, $\varepsilon =0$ for a perfect mirror, $0<\varepsilon <1$ for common gray bodies such as silicon). The output power by radiation for a cantilever infinitesimal element $\dd x$ with $H \ll B $ is thus $2 \varepsilon \sigma B \dd x T^4$. Including this radiative power into the energy balance, as well as the radiation received from the surroundings at temperature $T_0$, the steady state heat equation \eqref{HeatEquation0} becomes 
\begin{align}\label{Ray}
	\dv{}{x}\left( \lambda (T)\dv{T}{x} \right)&=\frac{P_a}{B H} \delta_D(x-x_0) + \varepsilon \frac{2\sigma}{H} (T^4 -T_0^4). 
\end{align}
According to eq.~\eqref{Ray} we can roughly evaluate the radiative heat effect relatively to the heat conduction effect by computing a dimensionless number $F = 2\varepsilon \sigma T^3 L^2 / \lambda H$. For our silicon cantilevers, $H=\SI{2.5}{\micro m}$, $\varepsilon = 0.7$\cite{Note_1,timans_emissivity_1998}, around $T=\SI{1000}{K}$ we get $F \approx 0.25$. The expected radiative effects are thus to be considered at high temperature, and one needs to solve \eqref{Ray} in order to accurately determine the temperature profile of such cantilever.
 
In figure~\ref{FigThRadiationEffect}-a, we compute the temperature profiles $ \theta(x)$ solving eq.~\eqref{Ray} with the boundary conditions of eqs.~\eqref{BCThermal}, considering ($\varepsilon=0.7$) or not ($\varepsilon=0$) the radiative effect for three absorbed powers $P_a$. As expected, the radiative effect lowers the temperature rise. The corresponding normalized temperature profiles are displayed in figure~\ref{FigThRadiationEffect}-b: the radiative effect has a negligible effect on the functions $\Theta(\xn)$. In the inset of figure~\ref{FigThRadiationEffect}-a, we report the maximum temperature increase $\Delta T$ as a function of $P_a$ up to the melting point. This function becomes dependent of the cantilever geometry and does not have the same variation as the spatial profiles. There is no equivalence anymore between the power $P_a$ and the position $x$ seen previously when conduction is the only thermal process taken into account. For a cantilever such as $H=\SI{2.5}{\micro m}$, $B=\SI{30}{\micro m}$ $L=\SI{500}{\micro m}$, one needs to impose 16\% more power to reach the melting point than without radiative effect. This value drops to 4\% with $L=\SI{250}{\micro m}$. As predicted by the dimensionless number $F$, the radiative effect increases as $L^2$.

In conclusion, as long as the temperature elevation is such that $T \ll (\lambda H/2\sigma \epsilon L^2)^{1/3}$, corresponding to $F \ll 1$, the radiative heat transfer effect can be neglected. Otherwise, the radiative effect must be taken into account to accurately predict the temperature elevation even if the normalised temperature profile $\Theta(\xn)$ remains unperturbed at first order.

\section{Mechanical problem: thermally induced frequency shift}\label{ModelFreqShift}

Let us describe the rectangular cantilever dynamics in the Euler Bernoulli approximation: the flexural modes of the cantilever are supposed to be only perpendicular to its length and uniform across its width. The strain can thus be described solely by the transverse deflection $w(x,t)$. The kinetic energy density of the vibrating beam is $1/2\mu (\partial w/\partial t)^2 $, where $\mu$ denotes the linear mass density. The potential energy density is the strain energy stored elastically in the bent cantilever $1/2 EI (\partial^2 w/\partial x^2)^2$ with $E$ the Young's modulus and $I=BH^3/12$ the second moment of inertia, and potential energy due to an optional external load $f$. Integrating these quantities over the length $L$, one obtains the total kinetic and potential energy
\begin{subequations}
\label{Energies}
\begin{align}
K &= \frac{1}{2}\displaystyle\int_0^L \mu \left(\frac{\partial w}{\partial t}\right)^2 \dd x ,\\
U &= \displaystyle\int_0^L \left[ \frac{1}{2} E I \left(\frac{\partial^2 w}{\partial x^2}\right)^2 - f w\right] \dd x.
\end{align}
\end{subequations}
All quantities $\mu$, $E$ et $I$ and $f$ may depend on the position $x$. By deriving the Lagrange's equation for the Lagrangian $K-U$, one obtains the well-known Euler-Bernoulli equation governing the beam motion 
\begin{align}
	 \label{EBeq}
	 \frac{\partial^2}{\partial x^2} \left[ E I \frac{\partial^2w}{\partial x^2} \right] + \mu \frac{\partial^2 w}{\partial t^2} = f.
\end{align} 
The solution can be described as a Fourier expansion function
\begin{equation}
	w(x,t) = \sum_{n=1}^{\infty} W_n \phi_n(x) \cos(\omega_nt+\psi_n)
	\label{EqFourierExpDefl}
\end{equation}
where $\phi_n(x)$ are the normal modes and $\omega_n$ the natural frequencies. $W_n$ and $\psi_n$ are respectively the amplitude and the phase of the vibration mode. The normal modes $\phi_n(x)$ are the solutions of the time Fourier transform of eq.~\eqref{EBeq} in the absence of external force ($f=0$)
\begin{equation}
	 \label{EqBEmodal}
	 \frac{\dd^2}{\dd x^2} \left[ E I \frac{\dd ^2 \phi_n}{\dd x^2} \right] - \omega_n^2 \mu \phi_n =0,
\end{equation}
and must satisfy the appropriate boundary conditions. In our case of a clamped-free cantilever, these boundary conditions read as 
\begin{subequations}
\label{EqBC}
\begin{align}
\phi_n(0) & = 0, & \frac{ \dd \phi_n}{\dd x} (0) & = 0, \\
\frac{\dd^2 \phi_n}{ \dd x^2}(L) & = 0, & \frac{\dd^3 \phi_n}{ \dd x^3}(L) & = 0.
\end{align}
\end{subequations}
The normal modes function $\phi_n(x)$ form an orthonormal basis. It is possible to show that the total mechanical energy of the beam $K+U$ is the sum of independent terms corresponding to the mechanical energy of each mode \cite{butt_calculation_1995}. The cantilever can be seen as a collection of independent harmonic oscillators. Like any system free of non-conservative force, the mechanical energy associated to each mode $n$ remains constant in time. Inserting the modal vibration $W_n\phi_n(x)\cos(\omega_nt+\psi)$ into eqs.~\eqref{Energies}, the energy conservation allows us to get the frequencies of natural vibrations
\begin{align}
	\label{EqRD1}
	\omega_n^2 & = \frac{ \displaystyle\int_0^L E I \left. \phi_n''(x) \right. ^2 \dd x } { \displaystyle\int_0^L \mu \left. \phi_n(x) \right.^2 \dd x}, 
\end{align}
also known as the Rayleigh quotient \cite{rayleigh_theory_1877}. Knowing the function $\phi_n$ of various modes and substituting in \eqref{EqRD1}, the frequency of these modes of vibration can easily be calculated. It can be useful to rewrite eq.~\ref{EqRD1} using the normalised position $\xn=x/L$ and associated normal modes $\Phi_n(\xn)=\phi_n(L\xn)$:
\begin{align}
	\label{EqRD1normalised}
	\omega_n^2 & = \frac{ \displaystyle\int_0^1  E I \left. \Phi_n''(\xn) \right. ^2 \dd \xn} { L^4 \displaystyle\int_0^1  \mu \left. \Phi_n(\xn) \right.^2 \dd \xn}. 
\end{align}

In the case where the geometrical dimensions $B$, $H$ and the Young modulus $E$ remain constant along the length, eq.~\eqref{EqBEmodal} is reduced to a differential equation with constant coefficients. The exact forms of the normal modes $\Phi_n$ can be determined in terms of analytic functions:
\begin{equation} \label{EqModalfunction}
\begin{split}
\Phi_n(\xn) = & \cos \alpha_n \xn - \cosh \alpha_n \xn \\
& + \kappa_n ( \sin \alpha_n \xn- \sinh \alpha_n \xn )
\end{split}
\end{equation}
with $\kappa_n=(\cos \alpha_n + \cosh \alpha_n)/(\sin \alpha_n + \sinh \alpha_n)$.
The application of the boundary conditions \eqref{EqBC} only allows certain discrete values for the spatial eigenvalues $\alpha_n$. They are the roots of
\begin{equation}\label{EqSolbeta} 
	1+\cos \alpha_n \cosh \alpha_n =0,
\end{equation}
which leads to $\alpha_1=1.875$, $\alpha_2=4.694$, \ldots, and $\alpha_n=(n-1/2)\pi$ for large $n$.

Inserting the modal functions \eqref{EqModalfunction} into the Rayleigh quotient \eqref{EqRD1}, one obtains the resonant pulsations $\omega_{n}^0$ 
\begin{equation}
	\label{EqRDParamConst} 
	\omega_{n}^0 = \frac{\alpha_n^2}{L^2}\sqrt{\frac{EI}{\mu}}.
\end{equation}
For common materials, like silicon, rise in temperature simultaneously induces a material softening (decrease of $E$) and an increase of its dimensions due to thermal expansion. According to the dispersion equation \eqref{EqRDParamConst}, both effects will affect the resonance frequency. Around a reference temperature $T_0$, the thermal dependency of the Young's modulus can be described by its first order temperature coefficient
\begin{equation}
	\label{ECoeff}
	a_E(T_0) = \frac{1}{E_0} \left. \dv{E}{T} \right\vert_{T=T_0}.
\end{equation}
Similarly, the extent to which the cantilever material expands or retracts upon temperature change is expressed by the linear thermal expansion coefficient defined by
\begin{equation}
	\label{ExpansionCoeff}
	a_l(T_0) = \frac{1}{l_0} \left. \dv{l}{T} \right\vert_{T=T_0}.
\end{equation}
Both coefficients $a_E$ and $a_l$ are intrinsic functions of the constituting material and may vary with the reference temperature $T_0$. For silicon at \SI{298}{K}, $a_E\approx \SI{-64e-6}{K^{-1}}$ \cite{masolin_thermo-mechanical_2013, gysin_temperature_2004} and $a_l\approx \SI{2.5e-6}{K^{-1}}$ \cite{okada_precise_1984,okaji_absolute_1988,watanabe_linear_2004}. In the following, we describe quantitatively how the frequency $\omega_n$ evolves when the cantilever is submitted to the one-dimensional temperature profile $T(x)=T_0+\theta(x)$.

\subsection{Uniform temperature}

Let us consider the case where the temperature distribution is uniform, $\theta(x) = \Delta T$. In this case the dispersion equation \eqref{EqRDParamConst} remains valid, the resonance frequency can be rephrased as 
\begin{equation}
	\label{EqResonanceTempUni}
	\omega_{n}=\alpha_n^2 \sqrt{ \frac{ EBH^3}{12mL^3 } },	
\end{equation}
where the mass $m$ and $\alpha_n$ are the only constant terms upon the cantilever temperature change. Deriving Eq. \eqref{EqResonanceTempUni} with respect to the temperature around $T_0$, the frequency shift can be obtained as
\begin{equation}
	\label{FTUniform} 
	 \Delta \Omega_{n} = \frac{\omega_{n}-\omega_n^0} {\omega_n^0} = \frac{1} {2} \left( a_E +a_l \right) \Delta T.
\end{equation}
The cantilever relative frequency shift induced by a uniform temperature change is independent of the mode number $n$ and is equally proportional to the thermal coefficients of elasticity $a_E$ and expansion $a_l$. For silicon, the temperature effect on elasticity dominates over the thermal expansion, the overall coefficient $a_E+a_l$ is negative. A rise of temperature will tend to decrease the resonant frequency. In the derivation to get \eqref{FTUniform}, we assumed that $a_l$ is a scalar, independent of the direction. This assumption of isotropic thermal expansion is correct for silicon which is a crystal with a diamond-like structure~\cite{okada_precise_1984}. Note again that the coefficient $a_l$ and $a_E$ are defined at a specific temperature $T_0$ and may vary with it. 

\subsection{Arbitrary temperature profile}

In the general case where the temperature increase profile $\theta(x)$ imposed is not uniform, the material's elasticity $E$ as well as the cantilever cross section dimensions $B$ and $H$, thus its second moment of inertia $I$ and linear mass density $\mu$, are function of the variable $x$ through the temperature profile. Close to $T_0$,
\begin{subequations}
\label{EBHx}
\begin{align}
	E(x) &= E_0 (1+a_E \theta(x)), \\
	I(x) &= I_0 (1+4 a_l \theta(x)), \\
	\mu(x) &= \mu_0 (1-a_l \theta(x)). 
\end{align}
\end{subequations}
The cantilever length $L$ reads as
\begin{equation}
	L = L_0 + a_l \displaystyle\int_0^{L_0} \theta(x)\dd x.
\end{equation}

The determination of the corresponding normal mode functions $\phi_n$, leading to the resonance frequencies, implies to solve eq.~\eqref{EqBEmodal} with the varying parameters of eqs.~\eqref{EBHx}. For an arbitrary distribution $\theta(x)$, this differential equation does not have any analytical solution. However, for most materials, the thermo-induced mechanical effects are relatively small: $ \abs{a_E \theta (x) } \ll 1 $ and $\abs{a_l \theta (x) } \ll 1$. For silicon, even for the maximum possible temperature increase, from \SI{0}{K} to the melting temperature $T_m^{\mathrm{Si}}=\SI{1683}{K}$, one remains in this limit of small variations since $1/a_l \gg 1/a_E \approx \SI{15000}{K} \gg T_m^{\mathrm{Si}}$. Thus, the effect of temperature on the normal modes $\phi_n$ can be seen as a perturbation. In this limit, we show in the appendix~\ref{AppendixFreqShift} that the frequency shift is not sensitive to the normal mode variations at the first order. In the following we thus rely on eq.~\ref{EqModalfunction} for the expression of the normal modes.

From small variation of eq.~\ref{EqRD1normalised} around temperature $T_0$, the frequency shift induced by the temperature change $\theta (x)=\Theta(\xn)\Delta T$ can be expressed as a function of the known normal modes $\Phi_{n}(\xn)$ and reads as
 \begin{multline} 
	\label{FreqShiftNonUnif1} 
	 \Delta \Omega_n = \frac{1}{2} \left[ ( a_E+4 a_l) \displaystyle\int_0^1  \xn \Theta(\xn) p_n(\xn) \dd \xn + \right. \\
	 \left. a_l \displaystyle\int_0^1  \Theta(\xn) q_n(\xn) \dd \xn -4 a_l \displaystyle\int_0^1  \Theta (\xn) \dd \xn \right] \Delta T.
\end{multline}
The functions $p_n$ and $q_n$ are respectively the normalized square curvature and square amplitude
\begin{subequations}
\label{eqpq}
\begin{align}
	 p_n(\xn)&= \frac{\left. \Phi''_n (\xn) \right.^2}{ \displaystyle\int_0^1  \left. \Phi''_n (\xn) \right.^2 \dd \xn} = \frac{1}{\alpha_n^4} \left. \Phi''_n (\xn)\right.^2, \\
	 q_n(\xn)&= \frac{\left. \Phi_n (\xn) \right.^2}{ \displaystyle\int_0^1  \left. \Phi_n (\xn) \right.^2 \dd \xn } = \left. \Phi_n (\xn)\right.^2.
\end{align}
\end{subequations}
 The two first terms in \eqref{FreqShiftNonUnif1} involve the normal modes $\Phi_n$, these contributions are thus mode dependent. The first one involves the temperature profile weighted by the local curvature $p_n(\xn)$. It corresponds to the effect of bending energy change due to elasticity temperature dependency and transversal dilatation. The second one involves the temperature profile weighted by the local amplitude $q_n(\xn)$ and corresponds to the effect of the kinetic energy change caused by dilatation. The third and last term is independent of the mode number $n$ and corresponds to the effect of the longitudinal dilatation.

In eq~\eqref{FreqShiftNonUnif1} the coefficients $ a_E$ and $ a_l$ are assumed to be independent from the temperature. For temperature change $\Delta T$ relatively large, this assumption may not be longer possible. For instance, the silicon thermal expansion coefficient $a_l$ varies from \SI{2.5e-6}{K^{-1}} at room temperature to \SI{4.6e-6}{K^{-1}} at \SI{1500}{K}. In such cases, one needs to use the following generalized expression to describe the frequency shift
\begin{align} 
	 \Delta \Omega_n & = g_n(\Delta T) \notag \\
	 & = \frac{ 1}{2 } \displaystyle\int_0^1  \Big[ A_E\left(\Theta(\xn)\Delta T\right) p_n(\xn) + \notag \\ 
	 & \quad \quad \quad A_l\left(\Theta(\xn)\Delta T\right) (4p_n(\xn) + q_n(\xn) -4) \Big] \dd \xn, \label{FreqShiftNonUnif2} 
\end{align}
where the functions $A_E$ and $A_l$ are by definition
\begin{subequations}
\label{eqAEAl}
\begin{align}
	A_E(\theta) &= \int_{T_0}^{T_0+\theta}  a_E(T) \dd T = \frac{E( T_0+\theta)}{E_0} -1,\\
	A_l(\theta) &= \int_{T_0}^{T_0+\theta} a_l(T) \dd T. 
\end{align}
\end{subequations}
In the derivation \eqref{FreqShiftNonUnif2} nothing was assumed beyond the small elasticity change $ \abs{A_E(\theta(x)) } \ll1$, and small dimensions change $ \abs{A_l(\theta(x)) } \ll1$. 

\subsection{Frequency shift sensitivities to softening and dilatation}

In order to evaluate and compare the respective effects of the cantilever softening and its dilatation on the frequency shift it is useful to look at the first order coefficients in $a_E$ and $a_l$ of $\Delta \Omega_n$ given by \eqref{FreqShiftNonUnif1} 
 \begin{align} 
 	s_E=\frac{ \dd 	 \Delta \Omega_n }{ \dd a_E } & = \frac{ 1 }{2}P_n \Delta T, \\
	s_l=\frac{ \dd	 \Delta \Omega_n }{ \dd a_l } & = \frac{ 1 }{2}\left(4 P_n -4 \bar \Theta + Q_n \right) \Delta T, \label{SensitivityExpans} 
\end{align}
with
\begin{subequations}
\label{eqPQ}
\begin{align}
	 P_n& = \displaystyle\int_0^1  \Theta(\xn) p_n(\xn) \dd \xn, \\
	 Q_n & = \int_0^1  \Theta(\xn) q_n(\xn) \dd \xn,
\end{align}
\end{subequations}
and $ \bar \Theta = \int_0^1 \dd \xn \Theta $ the mean temperature profile. The coefficients $ P_n$ and $ Q_n$ correspond to the temperature profile projection on the square modal curvature and the square modal amplitude. The sensitivity to the elasticity temperature dependency $s_E$ only depends on the curvature coefficient $P_n$, whereas the sensitivity to the thermal expansion $s_l$ depends both on the curvature and amplitude coefficients and also of the mean temperature $ \bar \Theta $ which is independent of the mode number $n$. Note that the modal contributions in $P_n$ and $Q_n$ have a positive contribution, while the non-modal term is negative. 

In table~\ref{table::Coeff} we report the computed coefficients assuming constant ($ \Theta \equiv 1$), linear ($ \Theta \equiv \xn$) and quadratic ($ \Theta \equiv \xn^2$) temperature profile. For a uniform temperature, all $ P_n$, $Q_n$ and $ \bar \Theta $ equal unity, we thus retrieve the temperature sensitivity given by eq.~\eqref{FTUniform}, $s_E=s_l= \Delta T/2$. For large modes number, the curvature and amplitude tend to be uniformly distributed along the cantilever, thus the projection coefficients $P_n$, $ Q_n$ tend to the mean normalized temperature $ \bar \Theta $, namely 1/2 for the linear profile and 1/3 for the quadratic one. Therefore, the sensitivity ratio $s_l/s_E$ tends to unity: high mode number behave as if the temperature field is uniform at $\bar \Theta \Delta T$.

For low modes numbers (especially mode 1) the curvature is larger near the clamp, which is at lower temperature, thus $P_n<\bar\Theta$. The frequency shift of the first mode of a beam heated at its free end is thus less sensitive to the elasticity temperature dependency than when the beam temperature changes uniformly. On the contrary, mode 1 amplitude is larger close to the free end of the cantilever which is at higher temperature, thus $Q_1>\bar \Theta$, leading to a stronger sensitivity to the thermal expansion of the first mode with respect to a uniform temperature. For silicon, though the elastic effect is still dominant, neglecting the thermal expansion for mode 1 induces errors of order of 10\% for a linear temperature profile, and even close to 30\% for a quadratic profile. This shows the importance to take into account dilatation effects even for material with relatively small dilatation effect (like silicon) when one wants to determine accurately the temperature from the frequency shift, especially with the first resonance.

\begin{table}[!ht]
 \center
 \begin{tabular}{|c|c|ccccc|} 
 \hline
 \multicolumn{2}{|c|}{mode number $n$ } & 1 & 2 & 3 &4&$\infty$ \\
 \hline
 \hline
 \multirow{4}*{$\Theta(\xn)=1$}& $P_n$ & 0.5 & 0.5 & 0.5 & 0.5 & 0.5\\
 & $Q_n$ & 0.5 & 0.5 & 0.5 & 0.5 & 0.5\\ 
 & $s_l/s_E$ & 1 & 1 & 1 & 1 & 1\\
 & $s_la_l/s_Ea_E$ & -3.9 \% & -3.9 \%& -3.9 \% & -3.9 \% &- 3.9 \% \\ 
 \hline
 \hline
 \multirow{4}*{$\Theta(\xn)=\xn$}& $P_n$ & 0.193 & 0.406 & 0.468 & 0.483 & 0.5\\
 & $Q_n$ & 0.807 & 0.594 & 0.532 & 0.517 & 0.5\\ 
 & $s_l/s_E$ & -2.17 & 0.536 & 0.862 & 0.932 & 1\\
 & $s_la_l/s_Ea_E$ & 8.8 \% & -2.2 \%& -3.5 \% & -3.8 \% &- 3.9 \% \\ 
 \hline
 \hline
 \multirow{4}*{$\Theta(\xn)=\xn^2$}& $P_n$ & 0.062 & 0.225 & 0.295 & 0.313 & 0.333\\
 & $Q_n$ & 0.675 & 0.414 & 0.360 & 0.346 & 0.333\\
 & $s_l/s_E$ & -6.71 & -0.08 & 0.699 & 0.851 & 1\\
 & $s_la_l/s_Ea_E$ & 27 \% & 0.3 \%& -2.8 \% & -3.4 \% &- 3.9 \% \\
 \hline
 \end{tabular}
 \caption{\label{table::Coeff} Temperature modal projection coefficients $P_n$ and $Q_n$ computed for a constant $\Theta(\xn)=1$, linear $\Theta(\xn)=\xn$ and quadratic $\Theta(\xn)=\xn^2$ profile. The sensitivity to elasticity change $s_E$ and to thermal expansion $s_l$ depends on the mode number $n$ and the temperature profile $\Theta(\xn)$. The ratio $s_la_l/s_Ea_E$ is computed for silicon at \SI{22}{\degree C}.}
\end{table}

\subsection{Temperature increase from frequency shift for a silicon cantilever}\label{SectionGnForSilicon}

In section~\ref{sectionThermal}, the temperature profiles $\theta(x)=\Theta(\xn)\Delta T$ were determined for a cantilever heated by a laser at any position $x_0$ along its length. In the case where conduction is the only dissipating mechanism, and neglecting the temperature increase at the clamp, the functions $\Theta(\xn)$ are fully determined by conductivity $\lambda(T)$, the maximum temperature $\Delta T$ and the heating position $x_0$. $\Theta(\xn)$ is otherwise independent of the cantilever geometry, see eq.~\eqref{HeatEquation0}. In addition we have concluded that thermal radiation for a silicon cantilever have a negligible effect on these spatial profiles. 

In section \ref{ModelFreqShift}, we determined the frequency shift $\Delta \Omega_n$ as a function of the mechanical properties variation with temperature $a_l(T)$, $E(T)$ and the cantilever temperature profile $\Theta(\xn)\Delta T$. For a material whose mechanical properties are known such as silicon, it is thus possible to deduce the maximum temperature increase $\Delta T$:
\begin{align} \label{DTvsDFreqShift} 
	\Delta T&=g_n^{-1} \left( \Delta \Omega_n \right).
\end{align}

\begin{figure}[!htb]
 \centering
 \includegraphics[scale=.5]{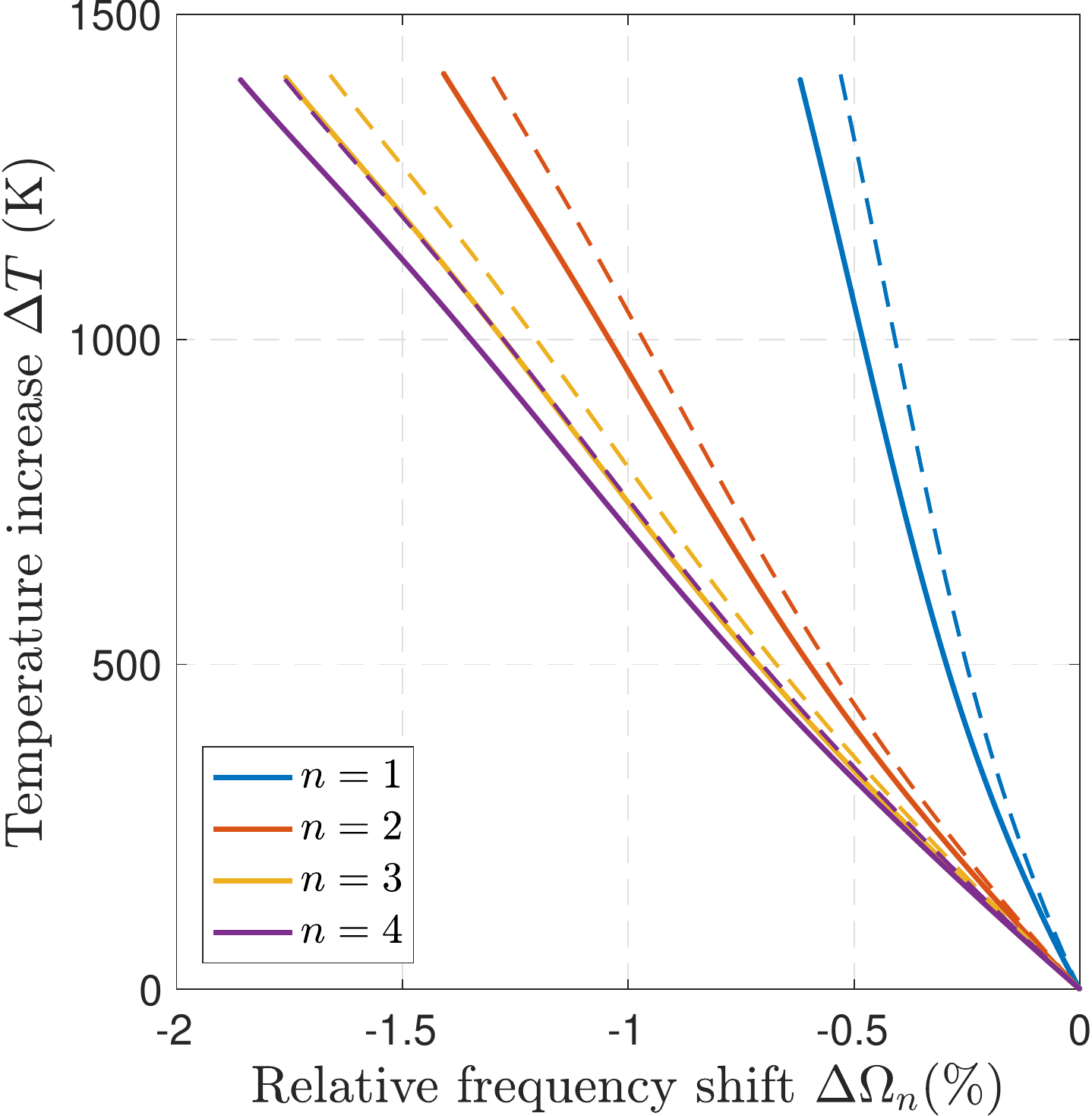}
 \caption{\label{CourbeShiftFreqvsT} Functions $g^{-1}_1$ to $g^{-1}_4$ computed for a silicon cantilever heated at its extremity in vacuum, supposing $l_{th}=L/20$ (plain line) or $l_{th}=0$ (dashed lines). These function are independent of the geometry and can be used for any silicon cantilever to infer $\Delta T$ (and thus the full temperature profile) from a measurement of $\Delta \Omega_n$.}
\end{figure}

The $g_n^{-1}$ functions are the inverse functions of $g_n$ which are defined according to eq.~\eqref{FreqShiftNonUnif2}. $g_n^{-1}$ functions can be obtained numerically by finding the value $\Delta T$ associated to any value of $\Delta \Omega_n$ such that $\Delta \Omega_n = g_n(\Delta T)$. In figure~\ref{CourbeShiftFreqvsT}, we computed the $g_n^{-1}$ functions for a silicon material up to the fourth mode when $x_0=L$. The temperature profiles $\Theta(\xn)$ are computed solving \eqref{EqTempSolveWithLth} such that $l_th=0$ or $l_\mathrm{th}/L=5\%$. This process can naturally be applied to any other choice of $x_0$ and $l_{th}$. We use the thermal conductivity displayed in figure~\ref{FigThProfileNoRadiation} from Ref.~ \onlinecite{glassbrenner_thermal_1964}. The Young modulus is described by the semi-empirical formula $E(T) = E_0 - B_ET \mathrm{exp} (-T_E^0/T) $ with the constants $E_0=\SI{167.5}{GPa}$, $B_E=\SI{15.8}{MPa/K}$, $T_E^0=\SI{317}{K}$ given by Ref.~\onlinecite{gysin_temperature_2004}. The thermal dilatation is described by the empirical formula given in Ref.~\onlinecite{okada_precise_1984}: $a_l(T)= C_1\left(1-\mathrm{exp} ( -C_2(T-C_3))+C_4T \right)\times 10^{-6}$ with $C_1=\SI{3.725}{K
^{-1}}$, $C_2=\SI{5.88}{K^{-1}}$, $C_3=\SI{124}{K}$ and $C_4=\SI{5.548E-4}{K^{-1}}$.

\section{Experimental validation}\label{Experiments}

\begin{figure}[htb]
 \centering
 \includegraphics[scale=.239]{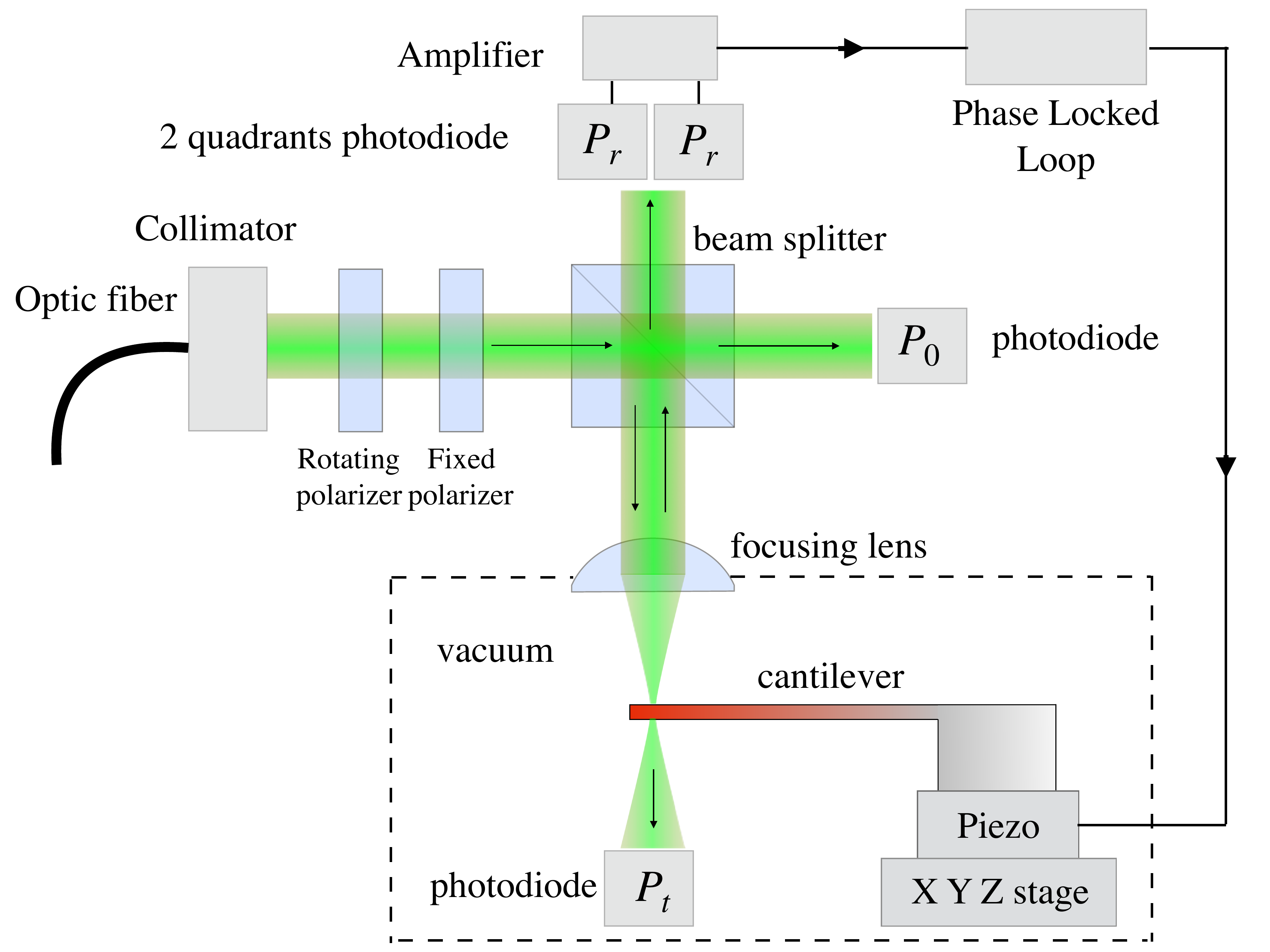}
 \caption{\label{FigSetup} Experimental setup to simultaneously heat and track the cantilever resonance frequencies. The cantilever placed in a vacuum chamber is illuminated by a \SI{532}{nm} laser beam, focused by a lens at its free end. The beam is partially absorbed, reflected, and transmitted by the cantilever. The reflected beam is sent on a two quadrants detector. The difference between the two quadrants is sent to a phase locked loop device, which drives the piezo element shaking the cantilever at resonance.}
\end{figure}

In order to test our approach, we perform resonance frequency measurements of silicon cantilevers heated in vacuum. In a previous paper \cite{aguilar_sandoval_resonance_2015}, we measured the frequency shift by detecting cantilever spontaneous fluctuations (thermal noise). Here we drive mechanically the cantilever and track its resonance frequencies with a phase locked loop (PLL). The advantage of such an active technique is to procure a much larger signal-to-noise ratio for a given acquisition time. Cantilever heating is induced by partial absorption of a laser beam focused at its free end. The fraction of the beam reflected off the cantilever is used as a sensing beam to detect the cantilever vibrations and allows us to track its resonance frequency shift (figure~\ref{FigSetup}). In the following, we detail the experiment and present the results obtained with two raw silicon cantilevers.

\subsection{Experimental setup}

We use a stabilized solid state laser from Spectra Physics, with a $\SI{40}{mW}$ maximum power available at \SI{532}{nm} to irradiate the cantilever. The incident beam power can be tuned continuously, benefiting from light polarization, by rotating two linear polarizers relatively to each other. The beam is then split in two: one beam is sent on a photodiode to measure the incident power $P_0$ while the other is focused through a lens at the cantilever free end. The laser spot size illuminating the cantilever is around $\SI{10}{\micro m}$ in diameter. The reflected beam is sent on a two-quadrant photodiode, measuring two signals $P_{r1}$ and $P_{r2}$. The difference signal $P_{r1}-P_{r2}$ delivered by the two quadrants is sensitive to the cantilever bending and used as the input signal of a PLL (Nanonis OC4). The output signal of the PLL drives a piezo actuator which vibrates the cantilever at its tracked resonance. The sum signal $P_{r1}+P_{r2}$ measures a fraction of the reflected power $P_r$ and allow us, after proper calibration, to know the cantilever reflectivity $R=P_r/P_0$. The power of the transmitted beam $P_t$ is measured by a photodiode placed under the cantilever and allows us to know the transmission coefficient $T=P_t/P_0$. The fraction $A$ of power absorbed by the cantilever is deduced from the two latter measurements: $A=1-R-T$. 

The cantilever is placed in a vacuum chamber at $\SI{1e-2}{mbar}$. At this pressure level, the contribution of convective heat transfer is negligible compared to thermal conduction \cite{lee_thermal_2007}.

We perform the experiments on two different geometries of cantilevers: cantilever C500 is $L=\SI{500}{\micro m}$ long (BudgetSensors AIO-CM) while cantilever C210 is $L=\SI{210}{\micro m}$ long (BudgetSensors AIO-FM). Both cantilevers have the same cross sections dimensions $B=\SI{30}{\micro m}$, $H=\SI{2.7}{\micro m}$, and are uncoated tipless atomic force microscope silicon cantilevers. Geometrical dimensions were measured using a scanning electron microscope (SEM) with uncertainties around 1\% for $L$ and $B$ and 4\% for $H$. For each tracked resonance, the incident power is continuously increased up to a maximal value then symmetrically decreased. The duration of one measurement is approximately 20 seconds. Since the characteristic time for heat diffusion ($\rho c_p L^2 /\lambda$) is respectively $\SI{3}{ms}$ and $\SI{0.5}{ms}$ for a $\SI{500}{\micro m}$ and $\SI{210}{\micro m}$ long cantilever, the temperature field for both cantilevers can safely be considered in the steady state regime during the whole experiment.

\subsection{Results: power scan at $x_0=L$}

\begin{figure*}[!tb]
 \centering
 \includegraphics[scale=.77]{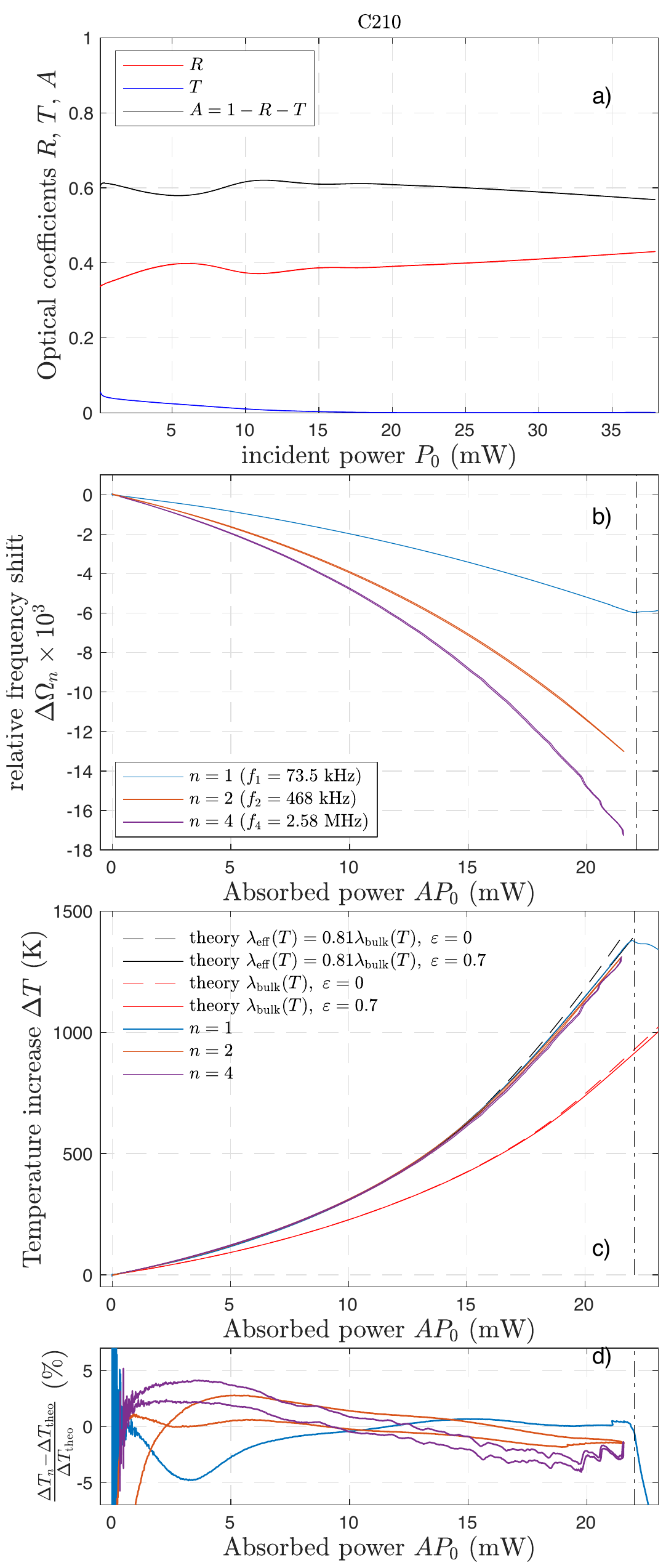}
 \hspace{4mm}
 \includegraphics[scale=.77]{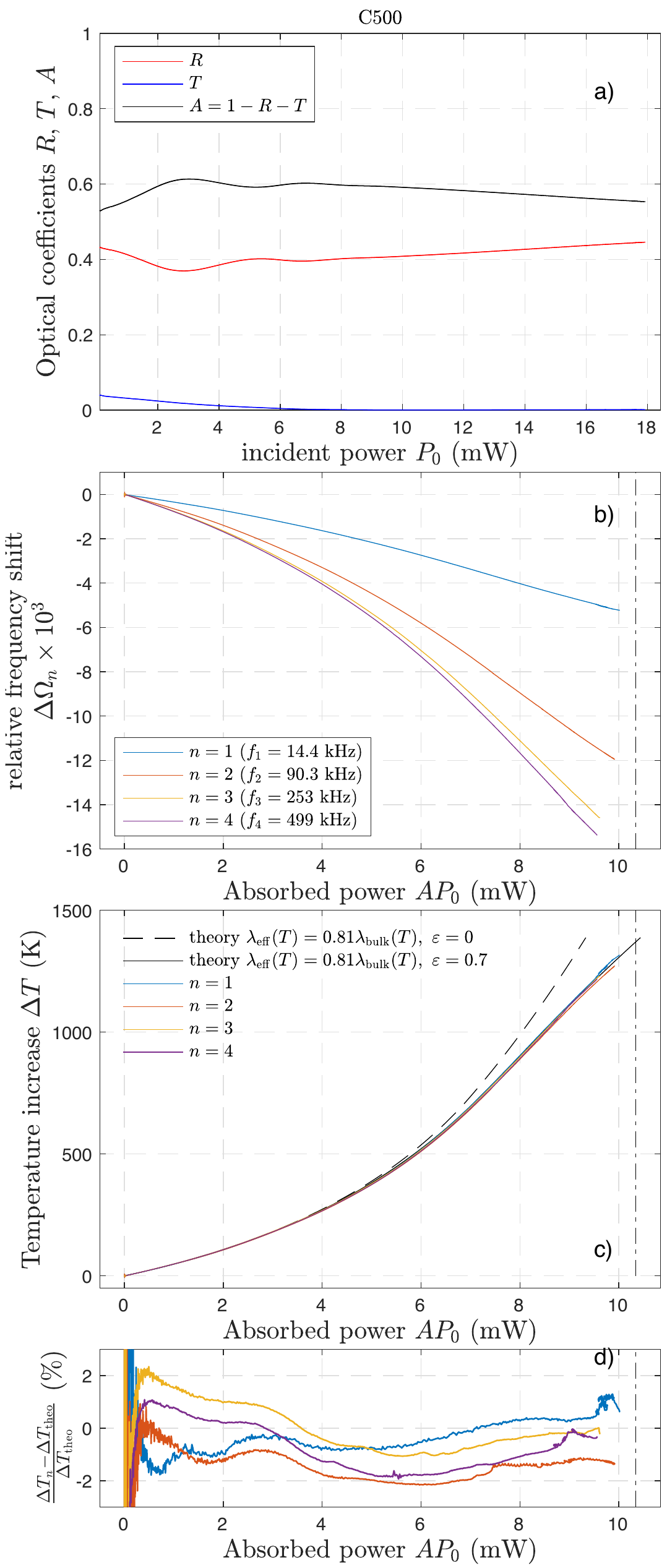}
 \caption{\label{FigC210-C500} Measured optical properties (a), frequency shift (b) and deduced temperature increase (c-d) for cantilever C210 (left) and C500 (right). a)~Measured optical reflectivity $R$, transmission $T$ and deduced absorption $A$ as a function of the incident beam $P_0$. The reflectivity and absorption oscillations can be explained by the temperature dependency of the silicon refractive index. b)~ Relative frequency shift $\Delta \Omega_n$ versus the absorbed power $AP_0$. During the mode 1 measurement, at $AP_0= \SI{22}{mW}$ for C210 and $AP_0= \SI{10.3}{mW}$ for C500 (dash-dot line), the cantilevers were molten at their extremity. c)~Temperature increase deduced from $\Delta \Omega_n$. All measurements are well superposed on a single curve. d)~Relative error between the temperature from frequency shift measurements $\Delta T_n$ and theory $\Delta T_\mathrm{theo}$ for $\lambda_ \mathrm{eff}(T)=0.81\lambda_\mathrm{bulk}(T)$ and $\varepsilon=0.7$. The agreement between all modes, and theory, is excellent.}
\end{figure*}

The measured optical reflectivity $R$, transmission $T$ and deduced absorption $A$ obtained with cantilever C210 and C500 are plotted in figure~\ref{FigC210-C500}-a. The optical reflectivity $R$ is calibrated by measuring the beam power reflected onto the chip, where the reflectivity is expected to be 37\% at \SI{532}{nm}~\cite{green_optical_1995}. We observe a variation of the optical coefficients with the incident optical power $P_0$. Indeed the silicon has an absorption depth of \SI{1.3}{\micro m} at \SI{532}{nm}, which is comparable to the cantilever thickness $H$. The cantilever is thus semi-transparent and has to be considered as a Fabry-Perot resonator with intrinsic optical losses. The reflected and transmitted light intensities result from the interferences between the reflections inside the cantilever. As the cantilever temperature increases, so does its refractive index, detuning the Fabry-Perot cavity. For the cantilever C210, the $\Delta T$ induced when $P_0$ increases from $0$ to $\SI{6}{mW}$ modifies interferences of the reflected beam from destructive to constructive, explaining the oscillations observed. For the cantilever C500, the reflected light interferes constructively at room temperature, thus this cantilever absorbs 15\% less power than the C210. This illustrates the importance to measure experimentally these coefficients if one wants to determine precisely the absorbed power.

The relative frequency shift $\Delta \Omega_n$ measured for the first four resonances\footnote{The third resonance of cantilever C210 couldn't be track efficiently with the PLL because of a spurious resonance of the piezo element in this frequency range} are displayed as a function of the absorbed power $AP_0$ in figure~\ref{FigC210-C500}-b. As expected for a cantilever in silicon, the temperature increase induces a red-shift of its frequencies resonances. This red-shift increases with the mode number $n$ and accelerate with the absorbed power. Note that for each measurement except $n=1$, we plot the frequency shift measured both for an increasing and decreasing power. The perfect superposition of the data indicates that the cantilever temperature is in the steady state regime during the whole experiment. Up to an absorbed power of $\SI{21.5}{mW}$ for C210 and $\SI{10}{mW}$ for C500, both optical coefficients and frequency shift measurements have an excellent reproducibility, the cantilever only undergoes reversible physical changes. During the measurement of the first mode, an additional power of \SI{1}{mW} is imposed leading to irreversible phenomena: the cantilever are molten at the beam spot. The absorbed power for melting the cantilevers ($\SI{22}{mW}$ and $\SI{10.3}{mW}$) differ by 47\%, which corresponds approximately to the cantilever length increase (42\%). Indeed, with conduction the main heat transfer mechanism, the cantilever temperature varies as the product $P_a L/BH$ (eq. \eqref{IntHeatEquation}).

In figure~\ref{FigC210-C500}-c, we report the temperature increase $\Delta T_n$ deduced from all measured relative frequency shift using $\Delta T_n=g_n^{-1}( \Delta \Omega_n)$ with $g_n$ given by eq.~\eqref{FreqShiftNonUnif2}. The temperature profiles $\Theta(x)$ used to compute $g_n^{-1}$ were obtained solving eq. \eqref{EqTempSolveWithLth} with $l_{th}= \SI{9}{ \micro m}$ from eq.~\eqref{ThermalLength} with $H=\SI{2.7}{ \micro m}$ and $l_\mathrm{chip}=\SI{300}{\micro m}$. We verify experimentally this value for $l_{th}$ with measurements where $x_0$ is set close to the clamp with a constant incident power $P_0$. The measured frequency shifts imply a temperature spatially varying as $x+l_{th}$, as suggested by our model. Because both cantilevers have identical cross section dimensions, the temperature at the clamp is expected to behave similarly. All temperatures $\Delta T_n$ deduced from the frequency shift are well superposed on the full range. At $AP_0= \SI{22}{mW}$ for C210 ($\SI{10.3}{mW}$ for C500), when the cantilever is damaged during the first mode measurement, the deduced temperature $\SI{1402}{ \degree C}$ ($\SI{1390}{ \degree C}$ for C500) is very close to the silicon melting temperature $T_m^{\mathrm{Si}}=\SI{1410}{\degree C}$. 

We can compare the deduced temperature $\Delta T_n$ to theoretical values obtained solving \eqref{Ray} imposing the boundary conditions \eqref{Tclamp} with $l_{th}= \SI{9}{\micro m}$. Using the bulk thermal conductivity values \cite{glassbrenner_thermal_1964}, the predicted temperature curve is significantly lower than the deduced temperatures. At the cantilever melting, the theory incorrectly predicts a temperature respectively 34\% and 27\% lower than the expected silicon melting temperature for C210 and C500. This discrepancy suggests that the effective cantilever conductivity is smaller than the bulk silicon conductivity. At micrometer scale, it has indeed been reported that the silicon conductivity is reduced due to phonon scattering at interfaces \cite{Johnson_direct_2013,minnich_thermal_2011}, deviating from the classical diffusion model. According to Ref.~\onlinecite{wang_computational_2014}, the conductivity of silicon film with a thickness $H=\SI{2.7}{\micro m}$ is expected to be approximately 20\% lower than the bulk value. Using an effective conductivity 19\% lower than the bulk silicon, we obtain an excellent agreement with the temperature deduced from the frequency shift, for both cantilevers. Note that the temperature profiles $\Theta(\xn)$ used to deduce the temperatures from the frequency shift only depend on the conductivity variation, see \eqref{HeatEquation0}. The deduced temperatures $\Delta T_n$ are thus independent of the choice of this multiplication factor which takes into account the phonon confinement. The relative error between the temperature increase from frequency shift measurements and the theory with $\lambda_ \mathrm{eff}(T)=0.81\lambda_\mathrm{bulk}(T)$ and $\varepsilon=0.7$ is below 5\% on the full temperature range for C210, and below 2\% for C500. We show the temperatures computed in both cases where the radiative heat transfer is considered ($\varepsilon=0.7$) or neglected ($\varepsilon=0$). For C210, the radiative effects appear at very high temperatures and are relatively small, but as expected from Fig.
~\ref{FigThRadiationEffect}, they are significant for C500.

\subsection{Results: position scan at fixed power} 
 
\begin{figure}[!htb]
 \centering
 \includegraphics{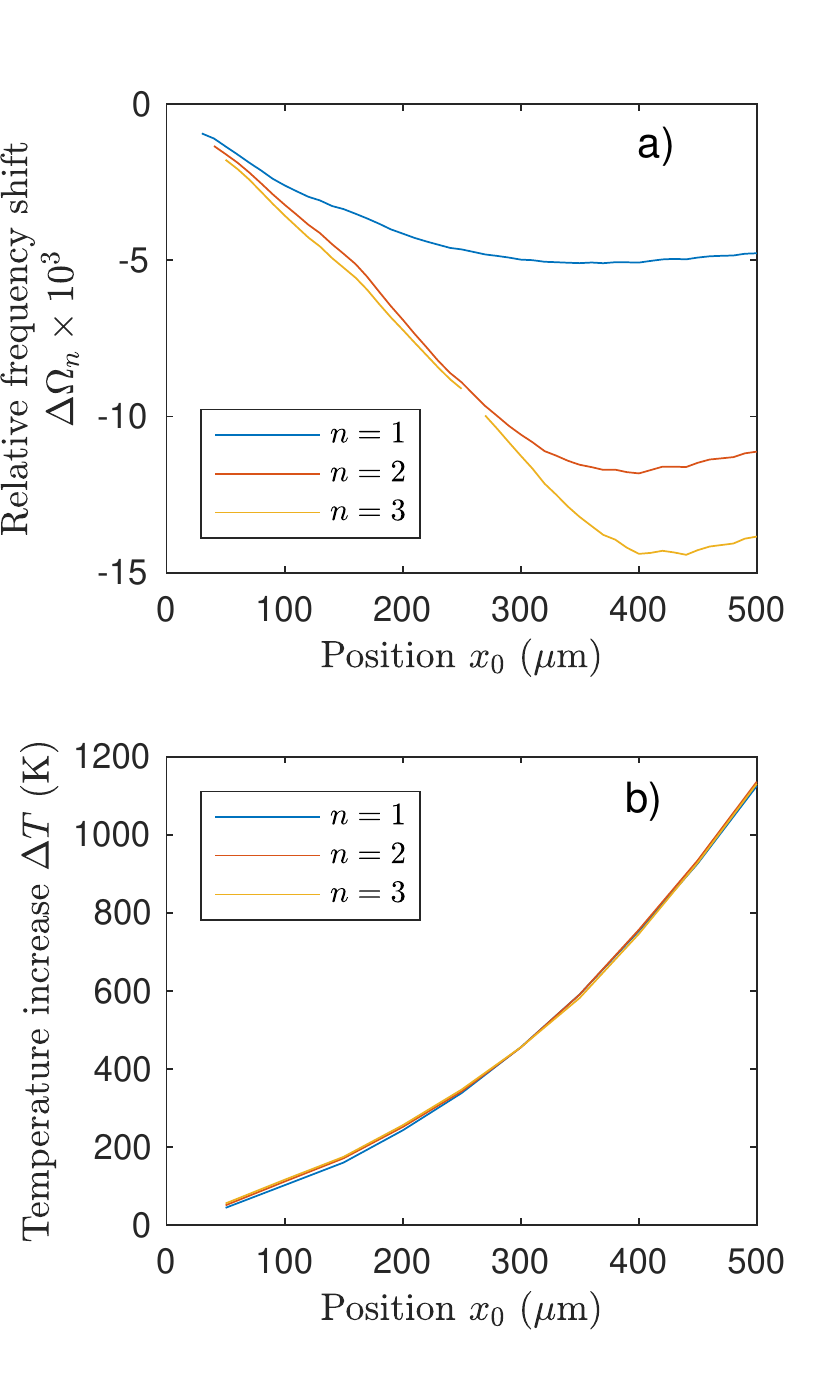}
 \caption{\label{FigFitC500} Cantilever C500: a) Frequency shifts measured by thermal noise when scanning in laser position $x_0$ at fixed laser power $P=\SI{13}{mW}$. b) Temperature increase $\Delta T$ deduced independently from the measured frequency shift using $\Delta T_n=g_n^{-1}(\Delta \Omega_n)$.}
\end{figure}

So far, the frequency shift has been measured as a function of the incident power illuminating the cantilever at its extremity $x_0=L$. For the cantilever C500, the frequency shift was also measured at fixed power as a function of the laser positions $x_0$ along the cantilever length (figure~\ref{FigFitC500}-a). Each tracked frequency shift were extracted from the thermal fluctuations spectrum measured as performed in Ref.~\onlinecite{aguilar_sandoval_resonance_2015}. When the laser beam is focused close to the clamped end acting as a cold thermostat, the heat flows only in a short portion of the cantilever, thus the warming is reduced (see figure~\ref{FigThProfileNoRadiationx0}). Accordingly, the measured frequency shifts for all three modes vanishes when pointing the laser at $x_0=0$. At the opposite, when pointing the laser far enough away from the clamped edge, the frequency shift becomes nearly independent from the laser position. This shows that the frequency shift is almost insensitive to the thermo-mechanical changes at the cantilever free end.

As in the previous paragraph, it is possible to deduce the temperature increase $\Delta T$ independently from every relative frequency shift using $\Delta T_n=g_n^{-1}(\Delta \Omega_n)$. In that case the functions $g_n^{-1}$ are computed using the temperature profiles displayed in figure~\ref{FigThProfileNoRadiationx0} described by \eqref{IntHeatEquation}. The deduced temperature increase $\Delta T_n$ as a function of the laser position $x_0$ are reported in figure~\ref{FigFitC500}-b and are well superimposed on a single curve for the full range $0<x_0<L$.

\subsection{Frequency shift measurement as a temperature sensor} 

\begin{figure}[htb]
 \centering
 \includegraphics{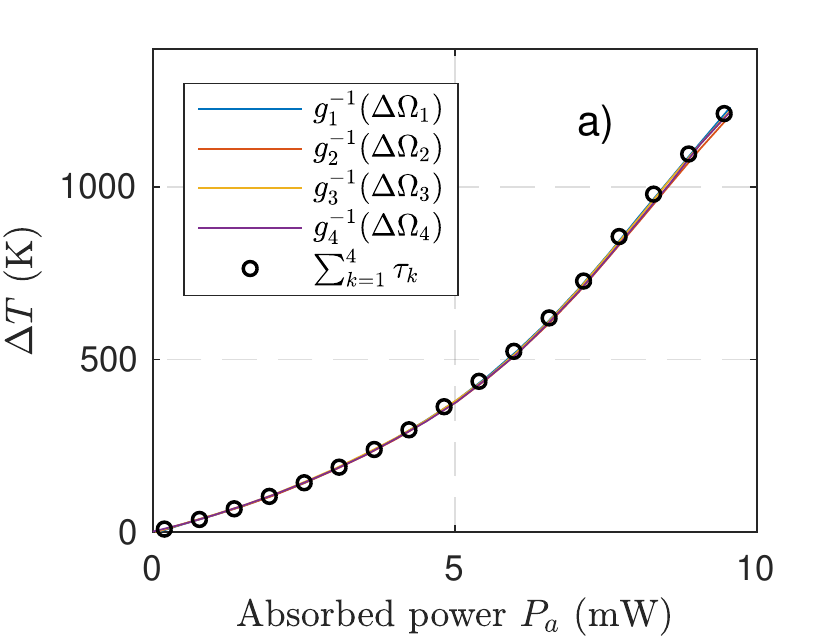}
 \includegraphics{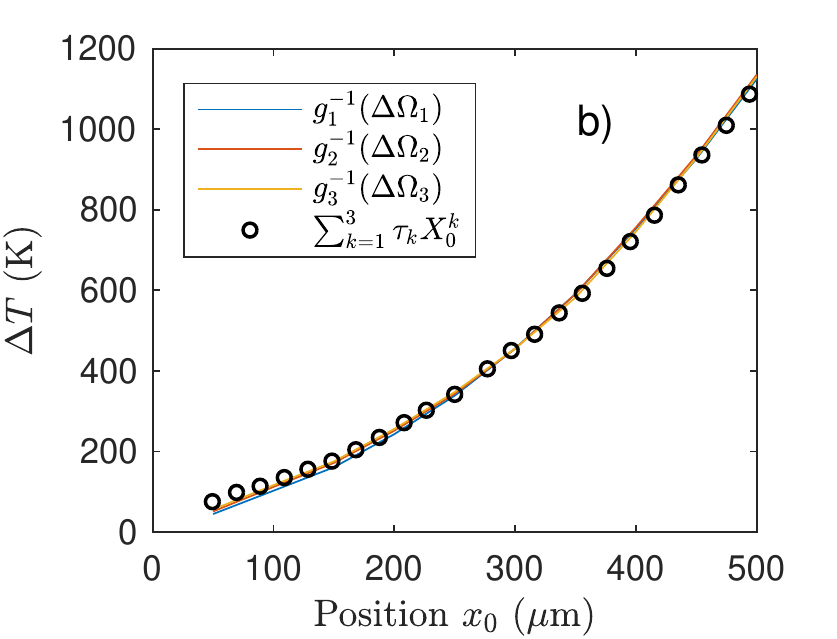}
 \caption{\label{FigFit} Cantilever C500: Temperature increase deduced from the frequency shifts using two different methods when varying the incident power (a) or scanning in laser position (b). The plain lines are computed for each mode with $\Delta T_n=g_n^{-1}(\Delta \Omega_n)$, assuming the temperature profile computed from the knowledge of the thermal conductivity (same data as figs.~\ref{FigC210-C500}-c and \ref{FigFitC500}-b). The circles correspond to the procedure described by eqs.~\ref{eq:thetapoly} and \ref{eq:DeltaOmegaTau}: at a given power or position, all frequency shifts are simultaneously used to guess the temperature profile, using an ad-hoc decomposition base, with no hypothesis on the thermal conductivity. Both methods are in very good accordance.}
\end{figure}

So far, the temperature increase $\Delta T_n$ has been deduced independently from each mode knowing \textit{a priori} the thermal conductivity $\lambda(T)$, thus the temperature distribution $\Theta(\xn)$. As a side-product of our study, we present here a different approach that takes advantage of the different frequency measurements and allows to retrieve $\Delta T$ only knowing the mechanical properties $E(T)$, $a_l(T)$. To do so, let us describe the temperature profile with a polynomial function of order $N$ such as
\begin{align} \label{eq:thetapoly}
	\theta(X) = \sum_{k=1}^N \tau_k X^k,
\end{align}
with thus $\Delta T = \sum_{k=1}^N \tau_k$. Using the theoretical expression \eqref{FreqShiftNonUnif2}, we get the expected values of the frequency shift as
\begin{align} 
	 \Delta \Omega_n & = \frac{1}{2} \int_0^1  \left[ A_E \left( \sum_{k=1}^N \tau_k X^k \right) p_n(\xn) + \right. \notag \\ 
	 & \left. \quad A_l\left(\sum_{k=1}^N \tau_k X^k\right) (4 p_n(\xn) + q_n(\xn) -4)\right] \dd \xn, \label{eq:DeltaOmegaTau}
\end{align}
with $p_n$, $q_n$, $A_l$ and $A_E$ defined by eqs.\ref{eqpq} and eqs.\ref{eqAEAl}. We therefore have a set of equations with $N$ unknowns ($\tau_{1},\ldots,\tau_{N}$). Measuring at least $N$ frequency shifts, we can thus perform a fit using a non linear least squared method. For each absorbed power $P_a$, we performed this fitting procedure with $N=4$ for the data of cantilever C500 displayed in figure~\ref{FigC210-C500}-b. The obtained temperature increase $\Delta T$ are reported in figure~\ref{FigFit}-a. The temperature deduced are in excellent agreement ($<0.8\%$ on average) with theory on the full temperature range.

The choice of a polynomial base is arbitrary, and can be tailored to any other information we have on the system. For example, if the heating position is at $\xn_0=x_0/L<1$, we can choose up to pad the polynomials above $\xn_0$ with a plateau. Using this description for the temperature, we performed the fitting procedure with $N=3$ for the data of cantilever C500 displayed figure~\ref{FigFitC500}-a. The obtained temperature increase $\Delta T$ are reported in figure~\ref{FigFit}-b. There is a good agreement ($<5\%$ on average) with the data deduced independently for each modes assuming the theoretical temperature profile.

With these two examples displayed on figure \ref{FigFit}, we demonstrate that at any fixed power or heating position, the measurement of a few frequency shifts is enough not only to measure the maximum temperature of the cantilever, but also to reconstruct the temperature profile along the cantilever. This last information could then be used to infer the cantilever thermal conductivity. 

\section{Conclusion} 

In this article, we propose a model that describes the shift of the mechanical resonance frequencies of a cantilever submitted to a temperature profile. We include both elasticity and geometry temperature dependency. While the elastic part makes the frequency shift sensitive to the temperature profile weighted by the local curvature, the thermal dilatation makes it also sensitive to region of large cantilever vibrations. The proposed model quantitatively describes the experimental data for a raw silicon cantilever heated locally in vacuum from room temperature up to the melting point. The thermal dilatation must be considered if one wants to deduce accurately the temperature from the frequency shift. This is particularly important for the first resonance mode: neglecting geometrical effects can lead to discrepancies up to 20\% in temperature estimation for a non uniform thermal gradient. In many experiments, the first oscillation mode of a resonator being the only measured one, such an effect should certainly be included to enable a quantitative measurement of the temperature profile.

The set of experiment provided, from Raman spectroscopy to frequency shifts measurements, demonstrates that our approach is pertinent for heating at any position on a cantilever: actually, scanning along its length proves to be a powerful tool to have a direct image of the temperature profile, thanks to a space - power equivalence in the heat equation when thermal radiation can be neglected.

As an opening, we demonstrate in the last section of the article that the temperature profile $\theta(x)$ can be evaluated thanks to the frequency shifts only, with no a priori knowledge of the thermal conductivity. The latter could then be extracted directly by differentiation of $\theta(x)$. The larger the number of resonance modes tracked, the best will be the approximation of the temperature profile. This approach only requires the knowledge of the temperature dependency of the elasticity and geometry. It could prove very interesting to evaluate the thermal conductivity of materials at the micro- and nano-scale, where confinement of phonons modifies significantly their properties from the bulk behavior. 

Though presented here for single-clamped cantilevers, the approach can be extended to other shapes of resonators, like double-clamped cantilever, tuning forks, membranes, or more complicate structures. One would only need to update the resonant mode shape and thermal profile to compute the frequency shifts, which can then be inverted to access the temperature field. 

\section*{Acknowledgement} 
F.M. acknowledges ANID-Chile through Fondecyt project N$^0$ 1201013 and Fondequip 130149. F.A. acknowledges CONICYT through Fondecyt Postdoctorado N$^0$ 3160695. Support from LIA-MSD France-Chile (Laboratoire International Associé CNRS, "Matière: Structure et Dynamique") is greatly acknowledged. Part of this research has been funded by the
ANR project HiResAFM (ANR-11-JS04-012-01) of the Agence Nationale de la Recherche in France.

\section*{Data availability}
The data that support the findings of this study are openly available in Zenodo at \url{https://doi.org/10.5281/zenodo.4629591}~\cite{Pottier-2021-Dataset-JAP}

\appendix

\section{Raman spectroscopy experiment} \label{AppendixRaman} 

In this appendix we describe the measurement of the temperature profile thanks to Raman spectroscopy, which results are plotted in Fig.~\ref{FigRaman}. The cantilever is placed in a closed cell with a glass optical window, vacuum pumped down to $\SI{7e-2}{mbar}$. Raman shifts are obtained through a commercial micro-raman confocal-spectrometer, Witec alpha 300, using a green laser at $\SI{532}{nm}$. The spot size on the cantilever is $\SI{10}{\micro m}$ in diameter, and can be moved under the microscope in an automated fashioned thanks to a motorised translation platform. The incident optical power $P_0$ is tuned by a pair of adjustable polarisers at the laser output, and measured before and after each scan in position with a powermeter. The sensor is placed in front of the cell window, and the measured value is corrected for the reflectivity of the glass window. $P_0$ is stable within $1\%$ during all the scans.

\begin{figure}[htb]
 \centering
 \includegraphics{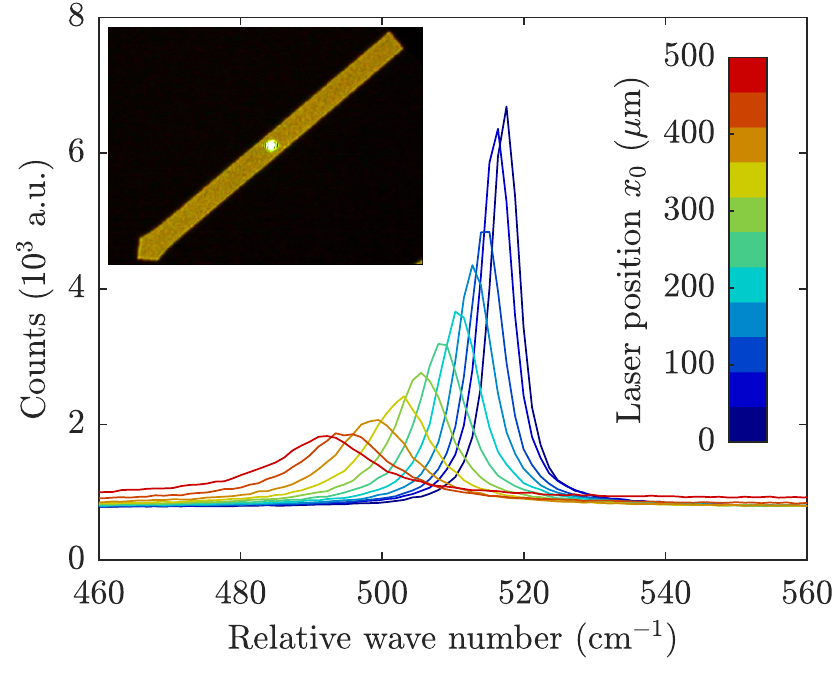} \caption{\label{FigRamanSpectra} Raman spectra acquired scanning position $x_0$ at constant incident power $P_0=\SI{15}{mW}$. Inset: microscope image of the cantilever with the laser spot in the center ($x_0=\SI{250}{\micro m}$).}
\end{figure}

In Fig.~\ref{FigRamanSpectra}, we plot the measured Raman spectra centered around a peak characteristic of silicon close to $\SI{520}{cm^{-1}}$, drifting and broadening when the temperature increases. Using the model for the temperature dependence of the Stoke's peak position of Ref.~\onlinecite{Balkanski-1983}, we can link the relative wavenumber at the peak to the temperature. Using the spectrometer pixel size, we would be limited to typically $\pm \SI{30}{K}$ of resolution. However we can use various strategies to estimate the peak position from the spectrum shape and not only the highest pixel, for instance by fitting the peak close to its maximum with a parabola. Using this approach, the estimate the resolution around $\pm \SI{10}{K}$ for the measured temperature of the tip.

It should be noted that the temperature dependence of the Stoke's peak position can be slightly different from the model of Ref.~\onlinecite{Balkanski-1983} in our case, since it as been reported that Raman shifts can depend on sample thickness or geometry~\cite{McCarthy-2005,Chen-2011}. As in Ref.~\onlinecite{Chen-2011} for instance, we had to offset the relative wavelengths by $\SI{4}{cm^{- 1}}$ to match the peak position at low power on the silicon chip with that expected at room temperature. The temperatures reported in Fig.~\ref{FigRaman} could thus need a correction in the $10-20\%$ range~\cite{McCarthy-2005,Chen-2011}. The conclusion that the temperature profile is driven by the product position times power is anyway robust to this minor calibration issue.

\section{Resonance frequency shift dependence on the modal shape} \label{AppendixFreqShift} 

When deriving eq.~\ref{FreqShiftNonUnif1}, we made the hypothesis that the modal shape changes due to the temperature field $\Theta(\xn) \Delta T$ could be ignored. In this appendix we demonstrate that they are indeed a second order effect. To this aim, let us expand the modal function to the first order in $\Delta T$: 
\begin{equation}\label{modalexpansion}
	\Phi_{n}(\xn)=\Phi_{n}^0(\xn) + a_\Phi \Delta T \Phi_{n}^1(\xn),
\end{equation}
with $a_\Phi \Delta T \ll 1$ and where $\Phi_{n}^0(\xn)$ is the modal function at uniform temperature $T_0$ given by eq.~\eqref{EqModalfunction} and $\Phi_{n}^1(\xn)$ corresponds to the modal shape modification due to the imposed temperature profile. Expanding eq.~\ref{EqRD1normalised} to the first order in $\Delta T$ leads to 
 \begin{multline} 
	\label{FreqShiftNonUnif1bis} 
	 \Delta \Omega_n = \frac{1}{2} \Big[ \displaystyle\int_0^1 \Theta(\xn) (a_E+4 a_l) p_n(\xn)\dd \xn + \\
	 \left. \displaystyle\int_0^1  \Theta(\xn) a_l \big( q_n(\xn) -4 \big) \dd \xn + 2 a_\Phi D_n \right] \Delta T.
\end{multline}
with 
 \begin{align}
 	 D_n &= \frac{ \int_0^1 {\Phi_n^0}''(\xn) {\Phi_n^1}''(\xn)\dd \xn }{ \int_0^1  \left. {\Phi_n^0}''(\xn) \right.^2\dd \xn } - 
	 \frac{ \int_0^1  \Phi_n^0(\xn) \Phi_n^1(\xn) \dd \xn}{ \int_0^1  \left.\Phi_n^0 (\xn) \right.^2 \dd \xn} \notag\\
	 &= \int_0^1  \left[\frac{{\Phi_n^0}''(\xn) {\Phi_n^1}''(\xn)}{\alpha_n^4} - \Phi_n^0(\xn) \Phi_n^1(\xn) \right]\dd \xn.
\end{align}

We recognise eq.~\ref{FreqShiftNonUnif1} in eq.~\ref{FreqShiftNonUnif1bis}, with only the $D_n$ extra term. Since each functions $\Phi_n^0(\xn)$ and $\Phi_n^1(\xn)$ meet the boundary conditions \eqref{EqBC}, using 2 integration by parts we have
\begin{equation}
 \int_0^1  {\Phi_n^0}''(\xn) {\Phi_n^1}''(\xn) \dd \xn = \int_0^1  {\Phi_n^0}^{(4)}(\xn) \Phi_n^1 (\xn)\dd \xn
\end{equation}
And as $\Phi_n^0$ is solution of the Euler-Bernoulli equation, $\left. \Phi_n^0 \right.^{(4)} = \alpha_n^4 \Phi_n^0$, implying that $D=0$. To the first order in $\Delta T$, the frequency shifts $\Delta \Omega_n$ are thus insensitive to the modal shape modification by the temperature profile, and eq.~\ref{FreqShiftNonUnif1} can safely be applied using the unperturbed resonance modes. \vfill

\bibliography{ArticleShift}

\end{document}